\documentclass{MSI}

\usepackage{mathtools}
\usepackage{subcaption}

\usepackage{amsmath}
\usepackage{paralist}
\usepackage{graphics} 
\usepackage{epsfig} 
\usepackage{graphicx}  \usepackage{epstopdf}
\usepackage[colorlinks=true]{hyperref}
\hypersetup{urlcolor=blue, citecolor=red}

\def\currentvolume{X}
 \def\currentissue{X}
  \def\currentyear{200X}
   \def\currentmonth{XX}
    \def\ppages{X--XX}
     \def\DOI{10.3934/xx.xx.xx.xx}

\newtheorem{theorem}{Theorem}[section]
\newtheorem{corollary}{Corollary}

\newtheorem{lemma}[theorem]{Lemma}

\theoremstyle{definition}
\newtheorem{definition}[theorem]{Definition}
\newtheorem{remark}{Remark}

\title[Compound Hawkes Processes in LOBs] 
      {General Compound Hawkes Processes in Limit Order Books}

\author[Anatoliy Swishchuk and Aiden Huffman]{}


 \email{aiden.huffman@ucalgary.ca}
 \email{aswish@ucalgary.ca}

\thanks{The first and second authors are supported by NSERC}

\thanks{$^*$ Corresponding author: Aiden Huffman}

\DeclarePairedDelimiter\floor{\lfloor}{\rfloor}

\begin{document}
\maketitle

\centerline{\scshape Anatoliy Swishchuk$^*$}
\medskip
{\footnotesize
 \centerline{2500 University Dr. NW}
   \centerline{Calgary, AB, Canada}
   \centerline{T2N 1N4}
} 

\medskip

\centerline{\scshape Aiden Huffman}
\medskip
{\footnotesize
 \centerline{2500 University Dr. NW}
   \centerline{Calgary, AB, Canada}
   \centerline{T2N 1N4}
}

\bigskip

 \centerline{(Communicated by Tony Ware)}

\begin{abstract}
In this paper, we study various new Hawkes processes. Specifically, we construct general compound Hawkes processes and investigate their properties in limit order books. With regards to these general compound Hawkes processes, we prove a Law of Large Numbers (LLN) and a Functional Central Limit Theorems (FCLT) for several specific variations. We apply several of these FCLTs to limit order books to study the link between price volatility and order flow, where the volatility in mid-price changes is expressed in terms of parameters describing the arrival rates and mid-price process.
\end{abstract}

\section{Introduction}

The Hawkes process (HP) is named after its creator Alan Hawkes (1971, 1974), \cite{27}, \cite{28}. The HP is a simple point process equipped with a self-exciting property, clustering effect and long run memory. Through its dependence on the history of the process, the HP captures the temporal and cross sectional dependence of the event arrival process as well as the 'self-exciting' property observed in our empirical data on limit order books. Self-exciting point processes have recently been applied to high frequency data for price changes \cite{54} or order arrival times \cite{55}. HPs have seen their application in many areas, like genetics (2010) \cite{11}, occurrence of crime (2010) \cite{50}, bank defaults \cite{51} and earthquakes \cite{52}.

Point processes gained a significant amount of attention in statistics during the 1950s and 1960s. Cox (1955) \cite{16} introduced the notion of a doubly stochastic process Poisson process (called the Cox process now) and Bartlett (1963) \cite{8} investigated statistical methods for point processes based on their power spectral densities. Lewis (1964) \cite{34} formulated a point process model (for computer power failure patters) which was a step in the direction of the HP. A nice introduction to the theory of point processes can be found in Daley et al. (1988) \cite{17}. The first type of point process in the context of market microstructure is the autoregressive conditional duration (ACD) model introduced by Engel et al. (1998) \cite{19}.

A recent application of HP is in financial analysis, in particular limit order books. In this paper, we study various new Hawkes processes, namely general compound Hawkes processes to model the price process in limit order books. We prove a Law of Larges Numbers (LLN) and a Function Central Limit Theorem (FCLT) for specific cases of these processes. Several of these FCLTs are applied to limit order books where we use asymptotic methods to study the link between price volatility and order flow in our models. The volatility of the price changes is expressed in terms of parameters describing the arrival rates and price changes. We also present some numerical examples. The general compound Hawkes process was first introduced in \cite{40} to model the risk process in insurance and studied in detail in \cite{41}. In the paper \cite{43} we obtained functional CLTs and LLNs for general compound Hawkes processes with dependent orders and regime-switching compound Hawkes processes.

Bowsher (2007) \cite{6} was the first one who applied the HP to financial data modelling. Cartea et al. (2011) \cite{9} applied HP to model market order arrivals. Filimonov and Sornette (2012) \cite{25}  and Filimonov et al. (2013) \cite{26} applied the HPs to estimate the percentage of price changes caused by endogenous self-generated activity rather than by the exogenous impact of news or novel information. Bauwens and Hautsch (2009) \cite{7} used a five dimensional HP to estimate multivariate volatility between five stocks, based on price intensities. Hewlett (2006) \cite{29} used the instantaneous jump in the intensity caused by the occurrence of an event to qualify the market impact of that event, taking into account the cascading effect of secondary events causing further events. Hewlett (2006) \cite{29} also used the Hawkes model to derive optimal pricing strategies for market makers and optimal trading strategies for investors given that the rational market makers have the historic trading data. Large (2007) \cite{32} applied a Hawkes model for the purpose of investigating market impact, with a specific interest in order book resiliency. Specifically, he considered limit orders, market orders and cancellations on both the buy and sell side, and further categorizes these events based on their level of aggression, resulting in a ten dimensional Hawkes process. Other econometric models based on marked point processes with stochastic intensity include autoregressive conditional intensity (ACI) models with the intensity  depending on its history. Hasbrouck (1999) \cite{30} introduced a multivariate point process to model the different events of an order book but did not parametrize the intensity. We note that Br\'emaud et al. (1996) \cite{4} generalized the HP to its nonlinear form. Also, a functional central limit theorem for nonlinear Hawkes processes was obtained in Zhu (2013) \cite{49}. The 'Hawkes diffusion model' introduced in Ait-Sahalia et al. (2013) \cite{1} attempted to extend previous models of stock prices to include financial contagion. Chavez-Demoulin et al. (2012) \cite{12} used Hawkes processes to model high-frequency financial data. An application of affine point processes to portfolio credit risk may be found in Errais et al. (2010) \cite{24}. Some applications of Hawkes processes to financial data are also given in Embrechts et al. (2011) \cite{23}.

Cohen et al. (2014) \cite{14} derived an explicit filter for Markov modulated Hawkes processes. Vinkovskaya (2014) \cite{47} considered a regime-switching Hawkes process to model its dependency on the bid-ask spread in limit order book. Regime-switching models for pricing of European and American options were considered in Buffington et al. (2000) \cite{2} and Buffington et al. (2002) \cite{3}, respectively. Semi-Markov processes were applied to limit order books in \cite{44} to model the mid-price. We also note that level-1 limit order books with time dependent arrival rates $\lambda(t)$ were studied in \cite{13}, including the asymptotic distribution of the price process.

The paper by Bacry et al. (2015) \cite{5} proposes an overview of the recent academic literature devoted to the applications of Hawkes processes in finance. It is a nice survey of applications of Hawkes processes in finance. In general, the main models in high-frequency finance can be divided into univariate models, price models, impact models, order-book models and some systemic risk models, models accounting for news, high-dimensional models and clustering with graph models. The book by Cartea et al. (2015) \cite{10} developed models for algorithmic trading such as methods for executing large orders, market making, trading pairs of collections of assets, and executing in the dark pool. This book also contains a link from which several datasets can be downloaded, along with \textbf{MATLAB} code to assist in experimentation with the data.

A detailed description of the mathematical theory of Hawkes processes is given in Liniger (2009) \cite{33}. The paper by Laub et al. (2015) \cite{35} provides background, introduces the field and historical development, and touches upon all major aspects of Hawkes processes. The results of the current paper were first annouced in in \cite{42}

The paper is organized as follows. A definition of a Hawkes process and description of its properties are given in Section 2. General compound Hawkes processes are described in Section 3. Law of Large Numbers (LLN) and Functional Central Limit Theorems (FCLT) for various general compound Hawkes processes, including non-linear, in limit order books are proved in Section 4. Section 5 contains a numerical exploration of the derived diffusion limits to several datasets and finally Section 6 concludes the paper.

\section{Definition of Hawkes Processes (HPs)}

In this section we give various definitions and some properties of Hawkes processes which can be found in the existing literature (see, e.g., \cite{27}, \cite{28}, \cite{23} and \cite{48}, to name a few). They include in particular one dimensional and non-linear Hawkes processes.

\begin{definition}[Counting Process]

A counting process is a stochastic process $N(t)$ with $t\geq 0$, where $N(t)$ takes positive integer values and satisfies $N(0) = 0$. It is almost surely finite and a right-continuous step function with increments of size $+1$.

Denote by $\mathcal{F}^N(t)$, $t\geq 0$, the history of the arrival up to time t, that is, $\mathcal{F}^N(t)$, $t\geq 0$, is a filtration (an increasing sequence of $\sigma$-algebras).

A counting process $N(t)$ can be interpreted as a cumulative count of the number of arrivals into a system up to the current time $t$. The counting process can also be characterized by the sequence of random arrival times $(T_1,\ T_2,\ \dots)$ at which the counting process $N(t)$ has jumped. The process defined by these arrival times is called a point process (see \cite{17}).
\end{definition}

\begin{definition}[Point Process]

If a sequence of random variables $(T_1,\ T_2,\ \dots)$, taking values in $[0, \infty)$ has $P(0\leq T_1\leq T_2\leq \dots) = 1$, and the number of points in a bounded region is almost surely finite, then $(T_1,\ T_2,\ \dots)$ is called a point process.
\end{definition}

\begin{definition}[Conditional Intensity Function]\label{DCIF}

Consider a counting process $N(t)$ with associated histories $\mathcal{F}^N(t)$, $t\geq 0$. If a non-negative function $\lambda(t)$ exists such that
\begin{equation}\label{CIF}
    \lambda(t) = \lim_{h\to 0} \frac{E[N(t+h)-N(t)\ | \ \mathcal{F}^N(t)]}{h}
\end{equation}
then it is called the conditional intensity function of $N(t)$ (see \cite{35}). We note that originally this function was called the hazard function (see \cite{16}).
\end{definition}

\begin{definition}[One-dimensional Hawkes Process]\label{1DHP}

The one-dimensional Hawkes process (see \cite{35}, \cite{28}) is a point process $N(t)$ which is characterized by its intensity $\lambda(t)$ with respect to its natural filtration:
\begin{equation}\label{1DCIF}
    \lambda(t) = \lambda + \int_{0}^{t}\mu(t-s)dN(s)
\end{equation}
where $\lambda > 0$, and the response function $\mu(t)$ is a positive function that satisfies $\int_{0}^\infty \mu(s)ds < 1$.
\end{definition}

The constant $\lambda$ is called the background intensity and the function $\mu(t)$ is sometimes called the excitation function. To avoid the trivial case of a homogeneous Poisson process, we assume $\mu(t)\neq 0$. Thus, the Hawkes process is a non-Markovian extension of the Poisson process.

With respect to the Definitions of $\lambda(t)$ in \ref{DCIF} and $N(t)$ in \ref{1DHP}, it follows that

\begin{equation*}
    P(N(t+h)-N(t) = m\ | \ \mathcal{F}^N(t)) = \begin{cases}
    \lambda(t)h + o(h) & m = 1\\
    o(h) & m > 1\\
    1-\lambda(t)h + o(h) & m = 0
    \end{cases}
\end{equation*}

The interpretation of Equation \eqref{1DCIF} is that the events occur according to an intensity with a background intensity $\lambda$ which increases by $\mu(0)$ at each new event, eventually decaying back to the background intensity value according to the evolution of the function $\mu(t)$. Choosing $\mu(0) > 0$ leads to a jolt in the intensity at each new event, and this feature is often called the self-exciting feature. In other words, if an arrival causes the conditional intensity function $\lambda(t)$ in Equations \eqref{CIF}-\eqref{1DCIF} to increase then the process is called self-exciting.

We would like to mention that the conditional intensity function $\lambda(t)$ in Equations \eqref{CIF}-\eqref{1DCIF} can be associated with the compensator $\Lambda(t)$ of the counting process $N(t)$, that is

\begin{equation}\label{compensator}
    \Lambda(t) = \int_0^t \lambda(s)ds
\end{equation}

We note that $\Lambda(t)$ is the unique non-decreasing, $\mathcal{F}^N(t)$, $t\geq 0$, predictable function, with $\Lambda(0) = 0$ such that

\begin{equation*}
    N(t) = M(t) + \Lambda(t) \text{    a.s.}
\end{equation*}

where $M(t)$ is an $\mathcal{F}^N(t)$, $t\geq 0$, local martingale (existence of which is guaranteed by the Doob-Meyer decomposition).

A common choice for the function $\mu(t)$ in Equation \eqref{1DCIF} is the one of exponential decay (see \cite{27})

\begin{equation}\label{exp-excit}
    \mu(t) = \alpha e^{-\beta t}
\end{equation}

where parameters $\alpha$, $\beta > 0$. In this case, the Hawkes process is called the Hawkes process with exponentially decaying intensity.

In the case of Equation \eqref{exp-excit}, Equation \eqref{1DCIF} becomes 

\begin{equation}\label{exp-CIF}
    \lambda(t) = \lambda + \int_0^t \alpha e^{-\beta(t-s)}dN(s)
\end{equation}

We note that in the case of Equation \eqref{exp-excit}, the process $(N(t), \lambda(t))$ is a continuous-time Markov process, which is not the case for a general choice of excitation function in Equation \eqref{CIF}.

With some intitial condition $\lambda(0) = \lambda_0$, the conditional intensity $\lambda(t)$ in Equation \eqref{exp-CIF} with exponential decay in Equation \eqref{exp-excit} satisfies the SDE

\begin{equation}
    d\lambda(t) = \beta(\lambda-\lambda(t))dt + \alpha dN(t),\medskip t\geq 0
\end{equation}
which can be solved using stochastic calculus as

\begin{equation}
    \lambda(t) = e^{-\beta t}(\lambda_0-\lambda) + \lambda + \int_{0}^{t}\alpha e^{-\beta(t-s)}dN(s),
\end{equation}
which is an extension of Equation \eqref{exp-CIF}.

Another choice for $\mu(t)$ is a power law function

\begin{equation}\label{power-CIF}
    \lambda(t) = \lambda + \int_{0}^{t} \frac{k}{(c+(t-s))^p}dN(s)
\end{equation}
with positive parameters $(c, k, p)$. This power law form for $\lambda(t)$ in Equation \eqref{power-CIF} was applied in the geological model called Omori's law, and used to predict the rate of aftershocks caused by an earthquake.

\begin{definition}[D-dimensional Hawkes Process]

The D-dimensional Hawkes process (see \cite{23}) is a point process $\vec{N}(t) = (N^i(t))_{i=1}^D$ which is characterized by its intensity vector $\vec{t} = (\lambda^i(t))_{i=1}^D$ such that:

\begin{equation}\label{D-CIF}
    \lambda^i(t) = \lambda^i + \int_{0}^t \mu^{ij}(t-s)dN^j(s)
\end{equation}

where $\lambda^i > 0$, and $M(t) = (\mu^{ij}(t))$ is a matrix-valued kernel such that:
\begin{enumerate}
    \item it is component-wise non-negative: $(\mu^{ij}(t)) \geq 0$ for each $1\leq i,j \leq D$
    \item it is component-wise $L^1$-integrable
\end{enumerate}
In matrix-convolution form, Equation \eqref{D-CIF} can be written as

\begin{equation}
    \vec{\lambda}(t) = \vec{\lambda} + M*d\vec{N}(t)
\end{equation}
where $\vec{\lambda}(t) = (\lambda^i)_{i=1}^D$.
\end{definition}

\begin{definition}[Non-linear Hawkes Process]

The non-linear Hawkes process (see, e.g., \cite{48}) is defined by the intensity function in the following form:

\begin{equation}\label{nonlin-CIF}
    \lambda(t) = h\bigg(\lambda + \int_0^t \mu(t-s)dN(s)\bigg)
\end{equation}
where $h(\cdot)$ is a non-linear function with support in $\mathbb{R}^+$. Typical examples for $h(\cdot)$ are $h(x) = \mathbf{1}_{x\in\mathbb{R}^+}$ and $h(x) = e^x$.
\end{definition}

\begin{remark}
Many other generalizations of Hawkes processes have been proposed. They include mixed diffusion-Hawkes models \cite{24}, Hawkes models with shot noise exogenous events \cite{18} and Hawkes processes with generation dependent kernels \cite{37}, to name a few.
\end{remark}

\section{Compound Hawkes Processes}

In this section we define non-linear compound Hawkes process with $N$-state dependent orders. We also consider special cases of this general compound Hawkes process.

\begin{definition}[Non-Linear Compound Hawkes Process with $n$-state Dependent Orders (NLCHPnSDO) in Limit Order Books]\label{NLCHPnSDO}

Consider the mid-price process $S_t$

\begin{equation}\label{nonlinprocess}
    S_t = S_0 + \sum_{i=1}^{N_t} a(X_k)
\end{equation}

where $X_k$ is a continuous time $n$-state Markov chain, $a(x)$ is a continuous and bounded function on the state space $X := \{1,\ 2,\ ...,\ n\}$, $N(t)$ is the non-linear Hawkes process (see, e.g., \cite{48} defined by the intensity function in the following form (see Equation \eqref{nonlin-CIF}):

\begin{equation*}
    \lambda(t) = h\bigg(\lambda + \int_0^t \mu(t-s)dN(s)\bigg)
\end{equation*}

where $h(\cdot)$ is a non-linear increasing function with support in $\mathbb{R}^+$. We note that in \cite{4} it was shown that if $h(\cdot)$ is $\alpha$-Lipschitz (see \cite{4}) such that $\alpha ||h||_{L^1} < 1$ then there exists a unique stationary and ergodic Hawkes process satisfying the dynamics of Equation \eqref{nonlin-CIF}. We shall refer to the process in Equation \eqref{nonlinprocess} as a \textit{Non-linear Compound Hawkes Process with $n$-State Dependent Orders} (NLCHPnSDO). 

\end{definition} 

This non-linear compound Hawkes process will be the foundation for our studies throughout this paper. In the following subsection we will introduce four specific examples, which will be used for our empirical investigations of the mid-price processes. 

\subsection{Special Cases of Compound Hawkes Processes in Limit Order Books}\label{S3S1}

\begin{definition}[General Compound Hawkes Process With N-state Dependent Orders (GCHPnSDO)]\label{GCHPnSDO}

Suppose that $X_k$ is an ergodic continuous-time Markov chain, independent of $N(t)$, with state space $X = \{1,\ 2,\ ...,\ n\}$, $N(t)$ is a one-dimensional Hawkes process defined in Definition \ref{1DHP} and $a(x)$ is any bounded and continuous function on $X$. We define the \textit{General Compound Hawkes process with N-state Dependent Orders} (GCHPnSDO) by the following process

\begin{equation}\label{gchpnsdo}
    S_t = S_0 + \sum_{i=1}^{N(t)}a(X_k)
\end{equation}
Note that this process can be recovered from Equation \eqref{nonlinprocess} by letting $h(x) = x$.

\end{definition}

\begin{definition}[General Compound Hawkes Process with Two-State Dependent Orders (GCHP2SDO)]\label{GCHP2SDO}

Suppose that $X_k$ is an ergodic continuous time Markov chain, independent of $N(t)$, with two states $\{1,\ 2\}$. Then Equation \eqref{gchpnsdo} becomes

\begin{equation}\label{gchp2sdo}
    S_t = S_0 + \sum_{i=1}^{N(t)}a(X_k)
\end{equation}

where $a(X_k)$ takes only the values $a(1)$ and $a(2)$. Of course we can view this as a special case of the n-state case, where $n = 2$. This model was used in \cite{45} for the mid-price process in limit order books with non-fixed tick $\delta$ and two-valued price changes.

\end{definition}

\begin{definition}[General Compound Hawkes Process with Dependent Orders (GCHPDO)]\label{GCHPDO}

Suppose that $X_k \in \{-\delta,\ \delta\}$ and that $a(x) = x$, then $S_t$ in Equation \eqref{gchpnsdo} becomes

\begin{equation}\label{gchpdo}
    S_t = S_0 + \sum_{i=1}^{N(t)} X_k
\end{equation}

This type of process can be a model for the mid-price in limit order books, where $\delta$ is a fixed tick size and $N(t)$ is the number of order arrivals up to time $t$. We shall call this process a \textit{General Compound Hawkes Process with Dependent Orders} (GCHPDO). This is a generalization of the previous process, obtained by letting $a(1) = -\delta$ and $a(2) = \delta$.

\end{definition}

Having defined several mid-price processes, we now prove diffusion limit theorems and LLNs for each price process in the following Section. These diffusion processes will be used for our exploration of the applicability of this model to real world limit order book data.

\section{Diffusion Limits and LLNs}

\subsection{Diffusion Limit and LLN for NLCHPnSDO}

We consider the mid-price process $S_t$ defined in Definition \ref{NLCHPnSDO}, namely 

\begin{equation*}\label{nonlinenlgchpnsdo}
    S_t = S_0 + \sum_{i=1}^{N_t} a(X_k)
\end{equation*}

where $X_k$ is a continuous time n-state Markov chain and a(x) is a continuous bounded function on the state space $X = \{1,\, 2,\ ...,\ n\}$. $N_t$ is the number of price changes up to time $t$, described by the non-linear Hawkes process given in Equation \eqref{nonlin-CIF}.

\begin{theorem}[Diffusion Limit for NLCHPnSDO]
Let $X_k$ be an ergodic Markov chain with $n$ states $\{1,\ 2,\ ...,\ n\}$ and with ergodic probabilities $(\pi^*_1,\ \pi^*_2,\ ...,\ \pi^*_n)$. Let also $S_t$ be as defined in Definition \ref{NLCHPnSDO}, then
\begin{equation}
    \frac{S_{nt}-N(nt)\cdot \hat{a}^*}{\sqrt{n}}\xrightarrow{n\to\infty} \hat{\sigma}^* \sqrt{E[N[0,1]]}W_t
\end{equation}
where $W_t$ is a standard Wiener process and $E[N[0,1]]$ is the mean of the number of arrivals on a unit interval under the stationary and ergodic measure. Furthermore
\begin{equation}\label{muhatnonlin}
    0 < \hat{\mu} := \int_{0}^\infty \mu(s)ds < 1\ \text{and}\medskip \int_{0}^\infty s\mu(s)ds < \infty
\end{equation}

\begin{align}\label{nonlinnequations}
\begin{split}
    (\hat{\sigma}^*)^2 :&= \sum_{i\in X}\pi_i^* v(i)\\
    v(i) &= b(i)^2 + \sum_{j\in X}(g(j)-g(i))^2 P(i,j)-2b(i)\sum_{j\in X}(g(j)-g(i))P(i,j)\\
    b &= (b(1),\ b(2), ...,\ b(n))'\\
    b(i) :&= a(X_i) - a^* := a(i) - a^*\\
    g :&= (P+\Pi^*-I)^{-1}b\\
    \hat{a}^* :&= \sum_{i\in X}\pi_i^*a(X_i)
\end{split}
\end{align}

P is the transition probability matrix for $X_k$, i.e. $P(i,j) = P(X_{k+1} = j \ | \ X_k = i)$. $\Pi^*$ denotes the matrix of stationary distributions of $P$ and $g(j)$ is the $j$th entry of $g$.

\begin{proof}

From Equation \eqref{gchpnsdo} we have
\begin{equation}\label{nlprocessproof}
S_{nt} = S_0 + \sum_{i=1}^{N_{nt}}a(X_k)
\end{equation}
and
\begin{equation}
    S_{nt} = S_0 + \sum_{i=1}^{N_{nt}} (a(X_k)-\hat{a}^*) + N(nt)\hat{a}^*
\end{equation}
therefore
\begin{equation}\label{nlexpanded}
    \frac{S_{nt} - N_{nt}\hat{a}^*}{\sqrt{n}} = \frac{S_0 + \sum_{i=1}^{N_{nt}}(a(X_k)-\hat{a}^*)}{\sqrt{n}}.
\end{equation}
As long as $\frac{S_0}{\sqrt{n}} \xrightarrow{n\to\infty} 0$, we need only find the limit for
\[\frac{a(X_k)-\hat{a}^*}{\sqrt{n}}\]
when $n\to +\infty$. Consider the following sums
\begin{equation}
    \hat{R}_n^* := \sum_{k=1}^n (a(X_k)-\hat{a}^*)
\end{equation}
and
\begin{equation}
    \hat{U}_n^*(t) := n^{-1/2}[(1-(nt-\floor{nt}))]\hat{R}_{\floor{nt}}^* + (nt-\floor{nt})\hat{R}_{\floor{nt}+1}^*]
\end{equation}
where $\floor{\cdot}$ is the floor function.
Following the martingale method from \cite{45}, we have the following weak convergence in the Skorokhod topology (see \cite{39}):
\begin{equation}
    \hat{U}_{n}^*(t) \xrightarrow{n\to\infty}\hat{\sigma}^*W(t)
\end{equation}
We note that the results from \cite{4} imply by the ergodic theorem that
\begin{equation}
    \frac{N_t}{t}\xrightarrow{t\to\infty} E[N[0,1]]
\end{equation}
or
\begin{equation}\label{nonlinlln}
    \frac{N_{nt}}{nt}\xrightarrow{n\to\infty} tE[N[0,1]].
\end{equation}
Using the change of time $t \to N_{nt}/n$, we find that
\begin{equation}
\hat{U}_n^*(N_{nt}/n) \xrightarrow{n\to\infty}\hat{\sigma}^*W(tE[N[0,1]])
\end{equation}
or
\begin{equation}
    \hat{U}_n^*(N_{nt}/n)\xrightarrow{n\to\infty}\hat{\sigma}^*\sqrt{E[N[0,1]]}W(t)
\end{equation}
The result now follows from Equations \eqref{nlprocessproof}-\eqref{nlexpanded}
\end{proof}

\end{theorem}

\begin{lemma}[LLN for NLCHPnSDO]
The process $S_{nt}$ in Equation \eqref{nlprocessproof} satisfies the following weak convergence in the Skorokhod topology (see \cite{40}):
\begin{equation}
    \frac{S_{nt}}{n}\xrightarrow{n\to\infty}\hat{a}^*E[N[0,1]]t
\end{equation}
where $\hat{a}^*$ is defined in Equation \eqref{nonlinnequations} respectively.

\begin{proof}
From Equation \eqref{nonlinprocess} we have
\begin{equation}
    \frac{S_{nt}}{n} = \frac{S_0}{n} + \sum_{i=1}^{N_{nt}} \frac{a(X_k)}{n}
\end{equation}
The first term goes to zero when $n\to\infty$. On the right hand side, with respect to the strong LLN for Markov chains (see, e.g. \cite{38})
\begin{equation}
    \frac{1}{n} \sum_{k=1}^{n} a(X_k)\xrightarrow{n\to\infty}\hat{a}^*
\end{equation}
Then taking into account Equation \eqref{nonlinlln}, we obtain
\begin{equation}
    \sum_{i=1}^{N_{nt}} \frac{a(X_k)}{n} = \frac{N_{nt}}{n}\frac{1}{N_{nt}}\sum_{i=1}^{N_{nt}}a(X_k) \xrightarrow{n\to\infty}\hat{a}^*E[N[0,1]]t
\end{equation}
from which the desired result follows.
\end{proof}
\end{lemma}

\subsection{Diffusion Limit and LLN for GCHPnSDO}

We consider here the mid-price process $S_t$ which was defined in Definition \ref{GCHPnSDO}, namely

\begin{equation}
    S_t = S_0 + \sum_{i=1}^{N(t)} a(X_k)
\end{equation}

where $X_k$ is a continuous time n-state Markov chain, a(x) is a continuous and bounded function on the state space $X = \{1,\ 2,\ ...,\ n\}$, and $N(t)$ is the number of price changes up to time t, described by a one-dimensional Hawkes process defined in Definition \ref{1DHP}. This can be interpreted as the case of non-fixed tick sizes, n-valued price changes and dependent orders.

\begin{theorem}[Diffusion limit for GCHPnSDO]\label{thmgchpnsdo}
Let $X_k$ be an ergodic Markov chain with $n$ states $\{1,\ 2,\ ...,\ n\}$ and with ergodic probabilities $(\pi^*_1,\ \pi^*_2,\ ...,\ \pi^*_n)$. Let also $S_t$ be as defined in Definition \ref{GCHPnSDO}, then
\begin{equation}\label{gchpnsdores}
    \frac{S_{nt} - N(nt)\hat{a}^*}{\sqrt{n}}\xrightarrow{n\to\infty}\hat{\sigma}^*\sqrt{\lambda/(1-\hat{\mu})}W(t)
\end{equation}
where W(t) is a standard Wiener process,
\begin{equation}\label{gchpnsdomuhat}
    0 < \hat{\mu} := \int_0^\infty \mu(s)ds < 1\ \text{and}\ \int_{0}^\infty s\mu(s)ds < \infty
\end{equation}
\begin{align}\label{gchpnsdoeq}
\begin{split}
    (\hat{\sigma}^*)^2 :&= \sum_{i\in X}\pi_i^* v(i)\\
    v(i) &= b(i)^2 + \sum_{j\in X}(g(j)-g(i))^2 P(i,j)-2b(i)\sum_{j\in X}(g(j)-g(i))P(i,j)\\
    b &= (b(1),\ b(2), ...,\ b(n))'\\
    b(i) :&= a(X_i) - a^* := a(i) - a^*\\
    g :&= (P+\Pi^*-I)^{-1}b\\
    \hat{a}^* :&= \sum_{i\in X}\pi_i^*a(X_i)
\end{split}
\end{align}
P is the transition probability matrix for $X_k$, i.e. $P(i,j) = P(X_{k+1} = j \ | \ X_k = i)$. $\Pi^*$ denotes the matrix of stationary distributions of $P$ and $g(j)$ is the $j$th entry of $g$.

\begin{proof}
As in the previous theorem we have that
\begin{equation}\label{gchpnsdoproofb}
    S_{nt} = S_0 + \sum_{i=1}^{N(nt)}a(X_k)
\end{equation}
and
\begin{equation}
    S_{nt} = S_0 + \sum_{i=1}^{N(nt)}(a(X_k) - \hat{a}^*) + N(nt)\hat{a}^*
\end{equation}
where $\hat{a}^*$ is as defined above.

Therefore, we can conclude that
\begin{equation}
    \frac{S_{nt} - N(nt)\hat{a}^*}{\sqrt{n}} = \frac{S_0 + \sum_{i=1}^{N(nt)}(a(X_k)-\hat{a}^*)}{\sqrt{n}}
\end{equation}
as long as $\frac{S_0}{\sqrt{n}}\xrightarrow{n\to\infty}0$, we need only find the limit for
\[\frac{\sum_{i=1}^{N(nt)}(a(X_k)-\hat{a}^*)}{\sqrt{n}}\]
as $n\to\infty$. We consider the following sums
\begin{equation}
    \hat{R}^*_n = \sum_{k=1}^n (a(X_k)-\hat{a}^*)
\end{equation}
and
\begin{equation}
    \hat{U}_n^*(t) := n^{-1/2}[(1-(nt-\floor{nt}))\hat{R}^*_{\floor{nt}} + (nt-\floor{nt})\hat{R}^*_{\floor{nt}+1}]
\end{equation}
where $\floor{\cdot}$ is the floor function.

Then following the martingale method from \cite{45}, we have the following weak convergence in the Skorokhod topology (see \cite{39})
\begin{equation}\label{changeoftime}
    \hat{U}_n^*\xrightarrow{n\to\infty} \hat{\sigma}^* W(t)
\end{equation}
where $\hat{\sigma}^*$ is as defined above.

We note again that with respect to the LLN for Hawkes processes (see, e.g., \cite{17}) we have
\begin{equation}
\frac{N(t)}{t}\xrightarrow{t\to\infty}\frac{\lambda}{1-\hat{\mu}}
\end{equation}
or
\begin{equation}\label{gchpnsdoe}
\frac{N(nt)}{n}\xrightarrow{n\to\infty}\frac{t\lambda}{1-\hat{\mu}}
\end{equation}
where $\hat{\mu}$ is as defined above. Then using the change of time defined before where $t\to N(nt)/n$ we can find from Equations \eqref{changeoftime}-\eqref{gchpnsdoe}:
\[\hat{U}_n^*\bigg(\frac{N(nt)}{n}\bigg) \xrightarrow{n\to\infty}\hat{\sigma}^*W\bigg(\frac{t\lambda}{1-\hat{\mu}}\bigg)\]
or
\[\hat{U}_n^*\bigg(\frac{N(nt)}{n}\bigg) \xrightarrow{n\to\infty}\hat{\sigma}^*\sqrt{\frac{\lambda}{1-\hat{\mu}}}W(t)\]
The result in Equation \eqref{gchpnsdores} now follows from Equations \eqref{gchpnsdoproofb}-\eqref{gchpnsdoe}
\end{proof}
\end{theorem}

\begin{lemma}[LLN for GCHPnSDO]\label{llngchpnsdo}
The process $S_{nt}$ defined in Definition \ref{GCHPnSDO}, satisfies the following weak convergence in the Skorokhod topology (see \cite{40})
\begin{equation}
\frac{S_{nt}}{n}\xrightarrow{n\to\infty} \hat{a}^* \frac{\lambda}{1-\hat{\mu}}t    
\end{equation}
where $\hat{\mu}$ and $\hat{a}^*$ are defined in Equations \eqref{gchpnsdomuhat} and \eqref{gchpnsdoeq} respectively.
\begin{proof}
From Equation \eqref{gchpnsdo} we have
\begin{equation}
    \frac{S_{nt}}{n} = \frac{S_0}{n} + \sum_{i=1}^{N(nt)}\frac{a(X_k)}{n}.
\end{equation}
The first term goes to zero when $n\to\infty$. On the right hand side, with respect to the strong LLN for Markov chains (see, e.g., \cite{38})
\begin{equation}
    \frac{1}{n} \sum_{k=1}^n a(X_K) \xrightarrow{n\to\infty} \hat{a}^*
\end{equation}
Then taking into account Equation \eqref{gchpnsdoe} we obtain
\begin{equation}
    \sum_{i=1}^{N(nt)}\frac{a(X_k)}{n} = \frac{N(nt)}{n}\frac{1}{N(nt)}\sum_{i=1}^{N(nt)}a(X_k) \xrightarrow{n\to\infty} \hat{a}^* \frac{\lambda}{1-\hat{\mu}}t
\end{equation}
from which the desired result follows.
\end{proof}
\end{lemma}

\subsection{Diffusion Limits and LLNs for Special Cases of GCHPnSDO}

Theorem \ref{thmgchpnsdo} can be reduced to some of the special cases we outlined previously in Subsection\ref{S3S1}. Specifically in Definitions \ref{GCHP2SDO} and \ref{GCHPDO}, we consider the case of a 2-state Markov chain for which we provide the diffusion limit and LLN result as Corollaries below.

\hfill

We begin by considering the mid-price process $S_t$ (GCHP2SD) which was defined in \ref{GCHP2SDO}, namely
\begin{equation}
    S_t = S_0 + \sum_{i=1}^{N(t)} a(X_k)
\end{equation}
where $X_k$ is a continuous-time 2-state Markov chain, a(x) is a continuous and bounded function on $X = \{1,\ 2\}$ and $N(t)$ is the number of price changes up to moment t, described by a one-dimensional Hawkes process defined in Definition \ref{1DHP}. This can be interpreted as the case of non-fixed tick sizes, two-valued price changes and dependent orders.

\begin{corollary}[Diffusion Limit for GCHP2SDO]
Let $X_k$ be an ergodic Markov chain with two states $\{1,\ 2\}$ and with ergodic probabilities $(\pi^*_1,\ \pi^*_2)$. Further, let $S_t$ be defined as in Definition \ref{GCHP2SDO}. Then

\begin{equation}
    \frac{S_{nt} - N(nt)a^*}{\sqrt{n}}\xrightarrow{n\to\infty} \sigma^*\sqrt{\lambda/(1-\hat{\mu})}W(t)    
\end{equation}
where $W(t)$ is standard Wiener process,
\begin{equation}\label{gchp2sdomuhat}
    0 < \hat{\mu} := \int_0^\infty \mu(s)ds < 1 \  \text{and} \ \int_0^\infty \mu(s)ds < \infty
\end{equation}
\begin{align}\label{gchp2sdoeq}
\begin{split}
    (\sigma^*)^2 :&= \pi_1^* + \pi_2^*a_2^2 + (\pi_1^*a_1+\pi_2^*a_2)[-2a_1\pi_1^*-2a_2\pi_2^*+(\pi_1^*a_1+\pi_2^*a_2)(\pi_1^*+\pi_2^*)]\\
    &+ \frac{\pi_1^*(1-p)+\pi_2^*(1-p')(a_1-a_2)^2}{(p+p'-2)^2}\\
    &+ 2(a_2-a_1)\bigg[\frac{\pi_2^*a_2(1-p')-\pi_1^*a_1(1-p)}{p+p'-2}\\
    &+ \frac{(\pi_1^*a_1+\pi_2^*a_2)(\pi_1^*-p\pi_1^*-\pi_2^*+p'\pi_2^*)}{p+p'-2}\bigg]\\
    a^* :&= a_1\pi_1^* + a_2\pi_2^*
\end{split}
\end{align}
where $(p,p')$ are the transition probabilities of the Markov chain.
\end{corollary}

\begin{corollary}[LLN for GCHP2SDO]
    The process $S_{nt}$ defined in Definition \ref{GCHP2SDO} satisfies the following weak convergence in the Skorokhod topology (see \cite{40})
    \begin{equation}
        \frac{S_{nt}}{n} \xrightarrow{n\to\infty} \hat{a}^* \frac{\lambda}{1-\hat{\mu}}t
    \end{equation}
    where $\hat{\mu}$ and $\hat{a}^*$ are defined in Equations \eqref{gchp2sdomuhat} and \eqref{gchp2sdoeq} respectively.
\end{corollary}

Now let us consider the process $S_t$ defined in Definition \ref{GCHPDO}, specifically
\begin{equation}
     S_t = S_0 + \sum_{i=1}^{N(t)} X_k
\end{equation}
where $X_k \in \{-\delta, \delta\}$ is a continuous-time 2-state Markov chain, $\delta$ is the fixed tick size, and $N(t)$ is the number of price changes up to moment t, described by a one-dimensional Hawkes process defined in Definition \ref{1DHP}. This case we have a fixed tick size, two-valued price changes and dependent orders.

\begin{corollary}[Diffusion Limit for CHPDO]
    Let $X_k$ be an ergodic Markov chain with two states $\{\delta,\ -\delta\}$ and with ergodic probabilities $(\pi^*,\ 1-\pi^*)$, then taking $S_t$ as defined in Definition \ref{GCHPDO} then
    \begin{equation}
        \frac{S_{nt}-N(nt) a^*}{\sqrt{n}} \xrightarrow{n\to\infty} \sigma\sqrt{\frac{\lambda}{1-\hat{\mu}}}W(t)
    \end{equation}
    where $W(t)$ is a standard Wiener process,
    \begin{equation}
        0 < \hat{\mu} := \int_0^\infty \mu(s)ds < 1\  \text{and}\medskip \int_0^\infty \mu(s)ds < \infty
    \end{equation}
    \begin{align}
        \begin{split}
            a^* :&= \delta (2\pi^* - 1)\\
            \sigma^2 :&= 4\delta^2 \bigg(\frac{1-p'+\pi^*(p'-p)}{(p+p'-2)^2}-\pi^*(1-\pi^*)\bigg)
        \end{split}
    \end{align}
    and $(p,p')$ are the transition probabilities of the Markov chain $X_k$. We note that $\lambda$ and $\mu(t)$ are defined in Equation \eqref{1DCIF}.
\end{corollary}

We note that the LLN for both GCHP2SDO and CHPDO is identitical to the result given in Lemma \ref{llngchpnsdo}, after some simplification.

\section{Empirical Results}










In order to test the validity of our models and determine which best fits empirical data, we have considered level 1 LOB data for Apple, Amazon, Google, Microsoft and Intel on June 21st, 2012 \cite{53}. We first verify that the data is reasonable for our model by checking it's liquidity, this is illustrated in Table \ref{table:liquidity}.

\begin{table}[htpb]
\centering
\begin{tabular}{c|c|c}

Ticker & Avg \# of Orders per Second & Price Changes in 1 Day\\
\hline
AAPL & 51 & 64,350\\
AMZN & 25 & 27,557\\
GOOG & 21 & 24,084\\
MSFT & 173 & 3,217\\
INTC & 176 & 4,060\\

\end{tabular}
\caption{\label{table:liquidity} Stock liquidity of AAPL, AMZN, GOOG, MSFT and INTC for June 21st, 2012.}
\end{table}

\begin{figure}[htbp]

\centering
\begin{subfigure}[t]{0.49\textwidth}
\captionsetup{labelformat=empty}

\caption{\textbf{AAPL}}
\includegraphics[width=\textwidth, trim = 0 0 0 30, clip]{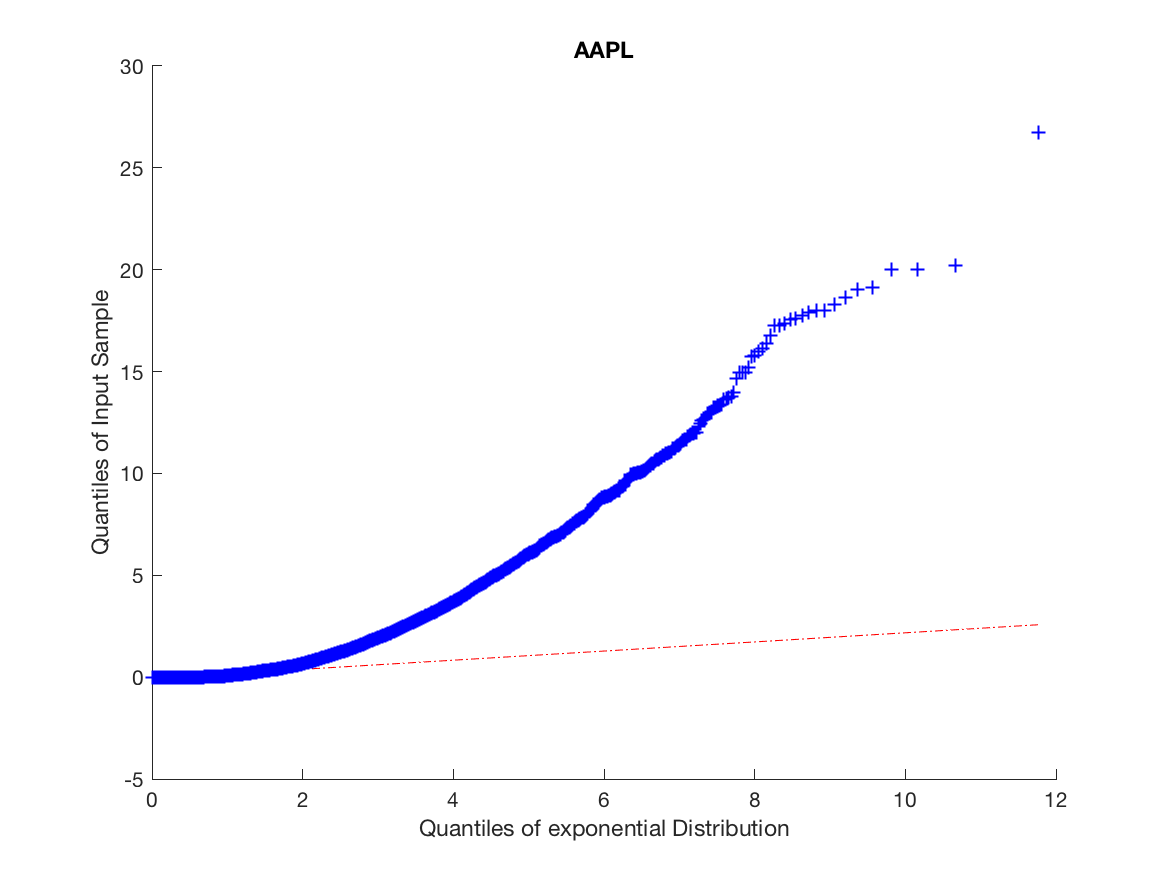}

\end{subfigure}
\begin{subfigure}[t]{0.49\textwidth}
\captionsetup{labelformat=empty}

\caption{\textbf{AMZN}}
\includegraphics[width=\textwidth, trim = 0 0 0 30, clip]{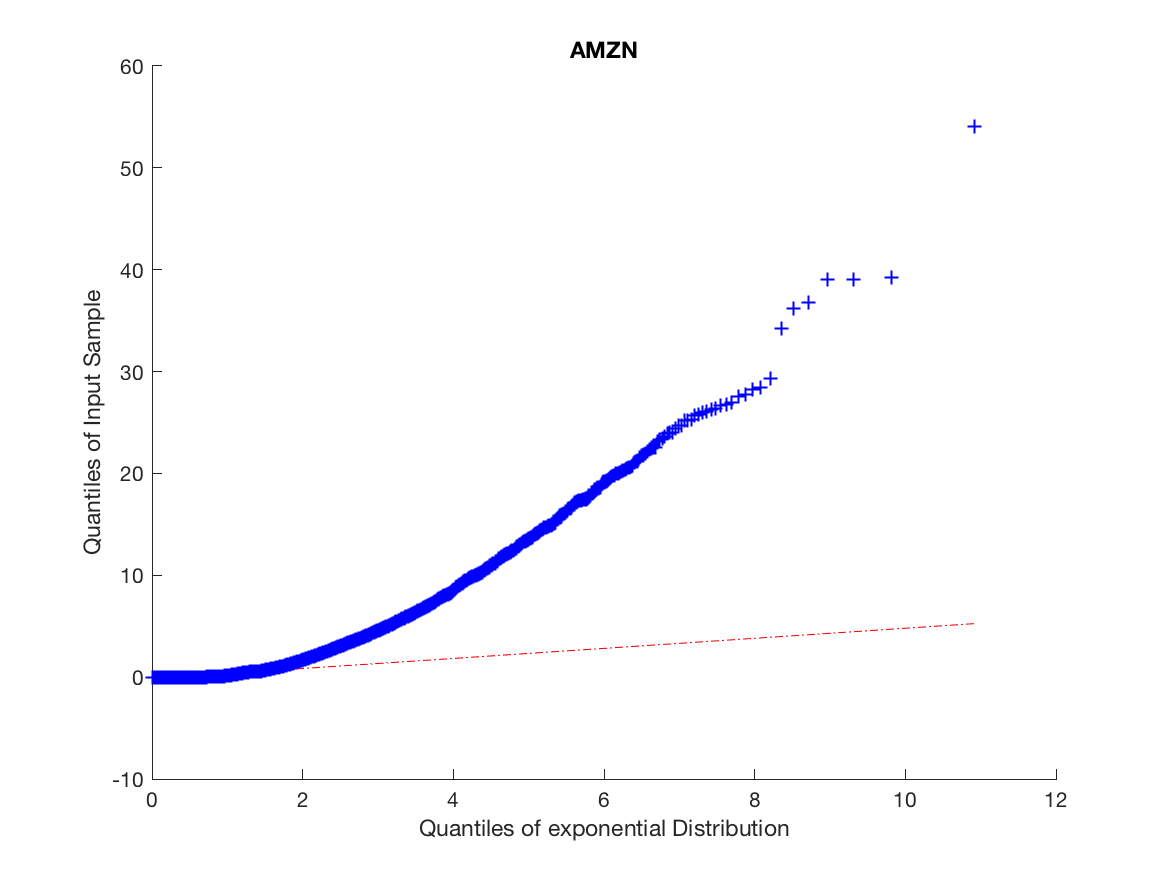}
\end{subfigure}

\vspace{3mm}

\begin{subfigure}[t]{0.49\textwidth}
\captionsetup{labelformat=empty}

\caption{\textbf{GOOG}}
\includegraphics[width=\textwidth, trim = 0 0 0 30, clip]{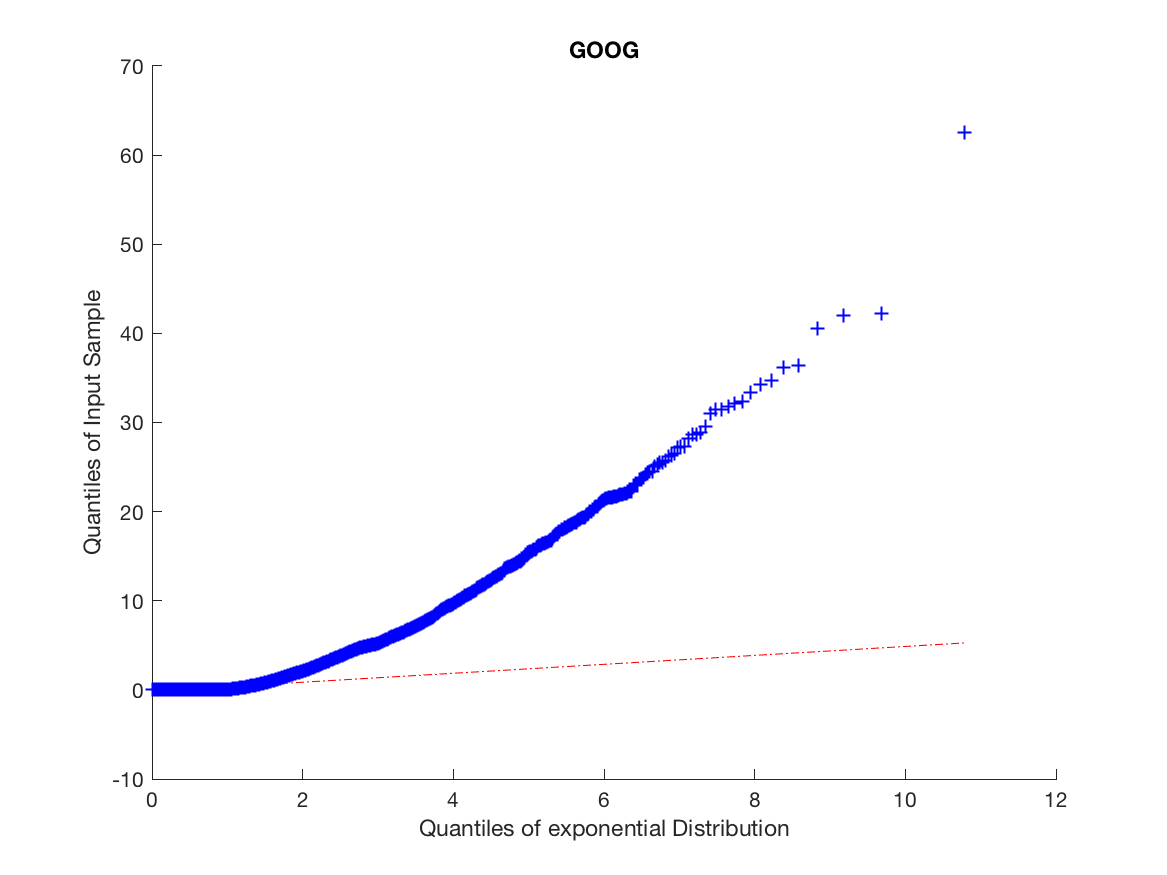}

\end{subfigure}
\begin{subfigure}[t]{0.49\textwidth}
\captionsetup{labelformat=empty}

\caption{\textbf{INTC}}
\includegraphics[width=\textwidth, trim = 0 0 0 30, clip]{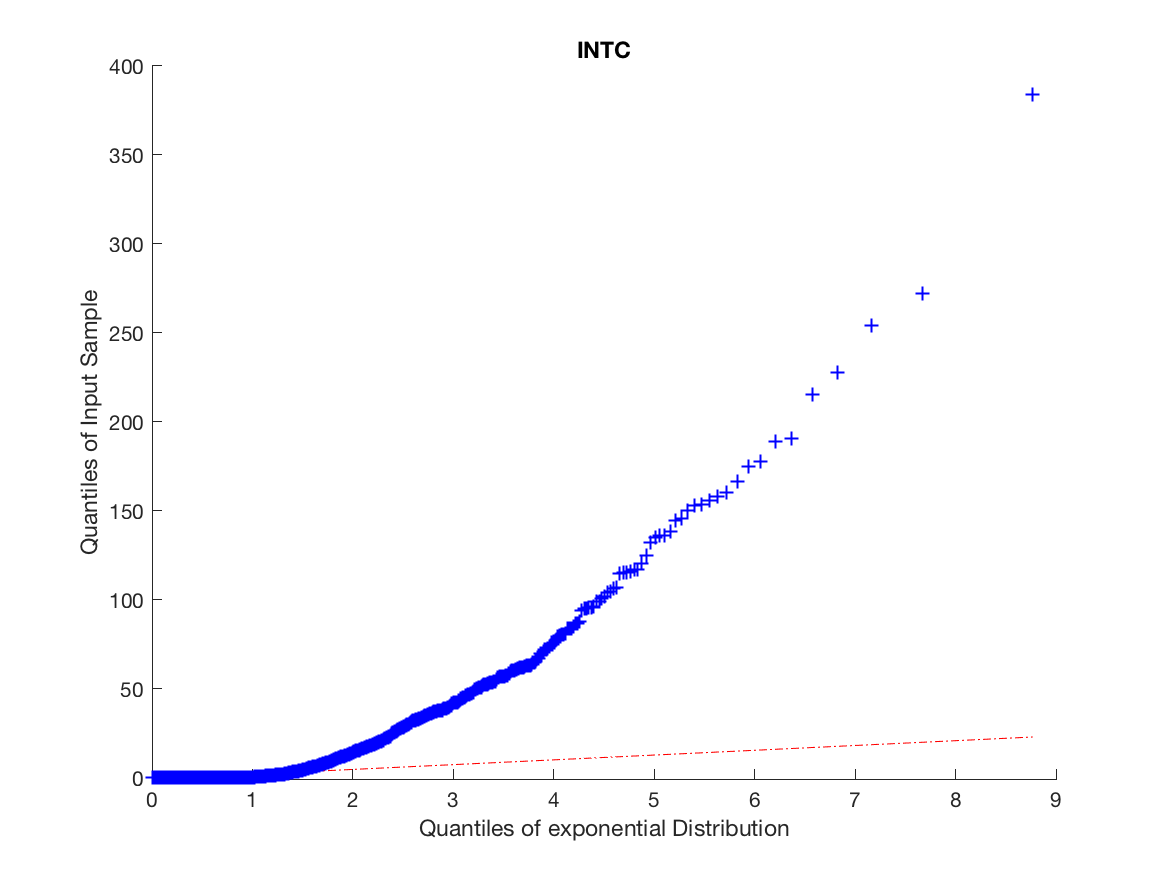}
\end{subfigure}

\vspace{3mm}

\begin{subfigure}[t]{0.49\textwidth}
\captionsetup{labelformat=empty}

\caption{\textbf{MSFT}}
\includegraphics[width=\textwidth, trim = 0 0 0 30, clip]{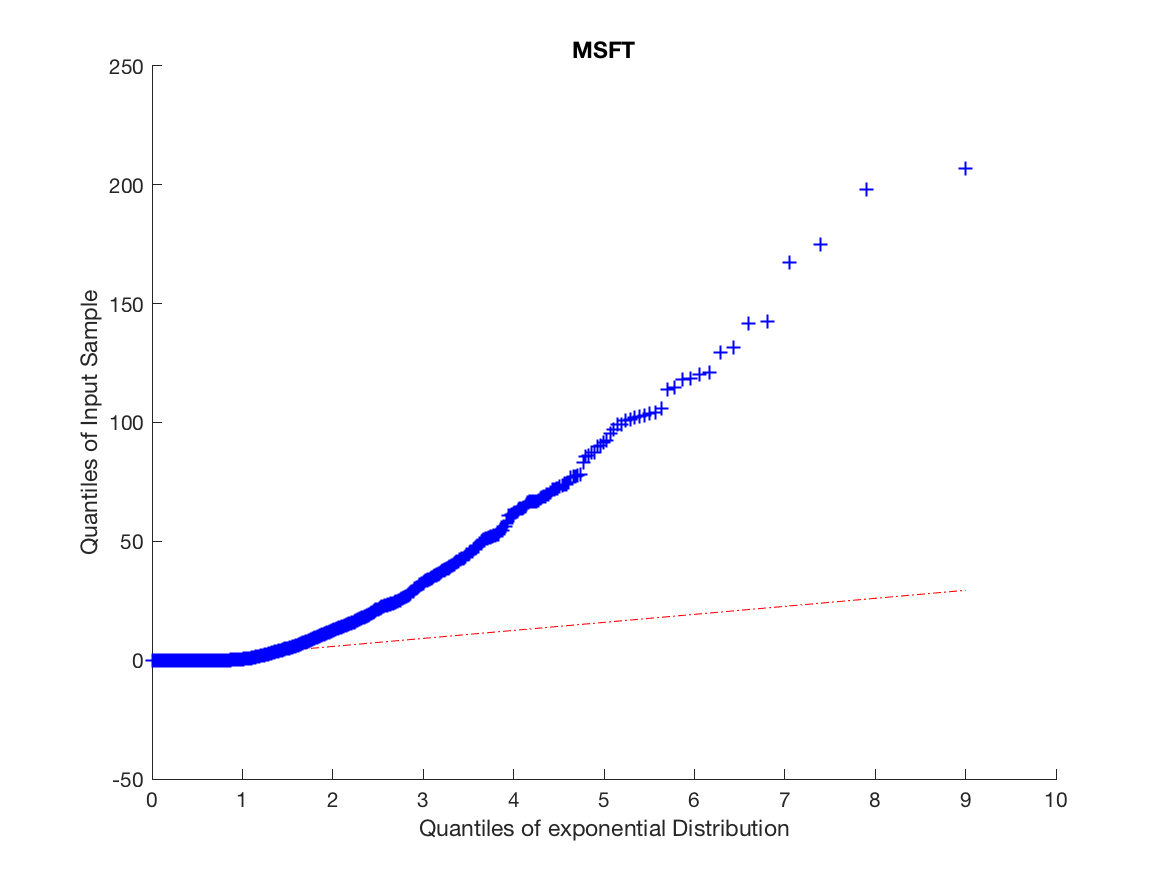}
\end{subfigure}

\caption{\label{fig:poisson} Above we provide a quantile-quantile plot of our empirical inter-arrival times against a Poisson process for each of the five stocks. We see that the inter-arrival data does not fit the expected curve, providing evidence that the underlying arrival process is not Poisson.}
\end{figure}

\begin{figure}[htbp]

\centering
\begin{subfigure}[t]{0.49\textwidth}
\captionsetup{labelformat=empty}

\caption{\textbf{AAPL}}
\includegraphics[width=\textwidth, trim = 0 0 0 30, clip]{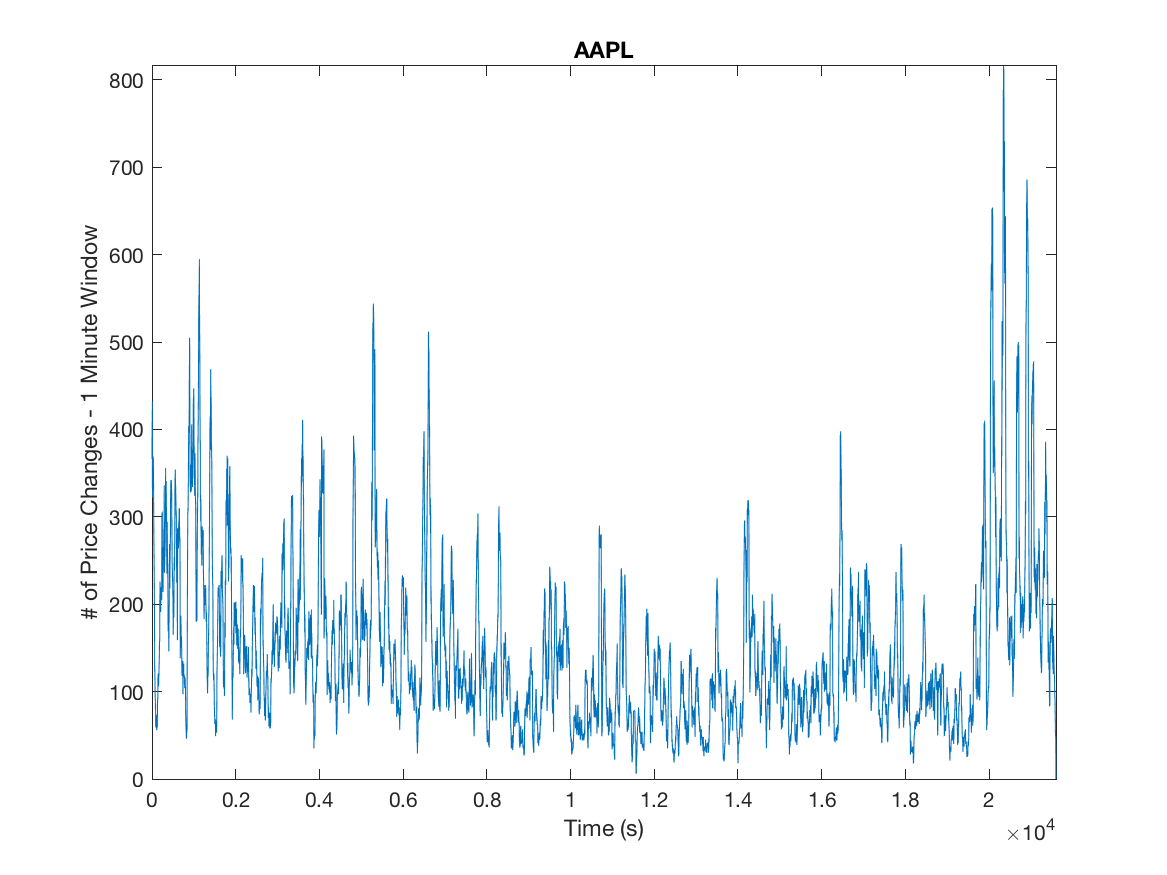}

\end{subfigure}
\begin{subfigure}[t]{0.49\textwidth}
\captionsetup{labelformat=empty}

\caption{\textbf{AMZN}}
\includegraphics[width=\textwidth, trim = 0 0 0 30, clip]{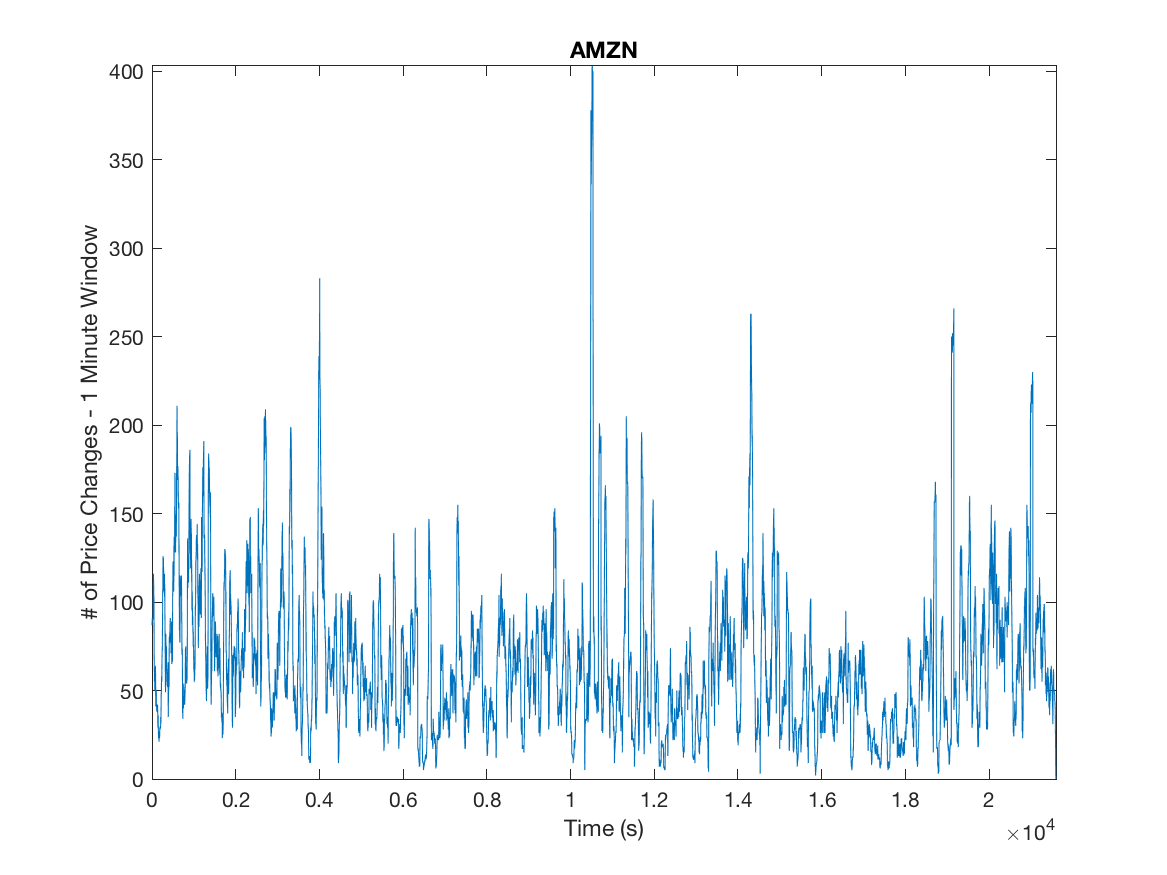}
\end{subfigure}

\vspace{3mm}

\begin{subfigure}[t]{0.49\textwidth}
\captionsetup{labelformat=empty}

\caption{\textbf{GOOG}}
\includegraphics[width=\textwidth, trim = 0 0 0 30, clip]{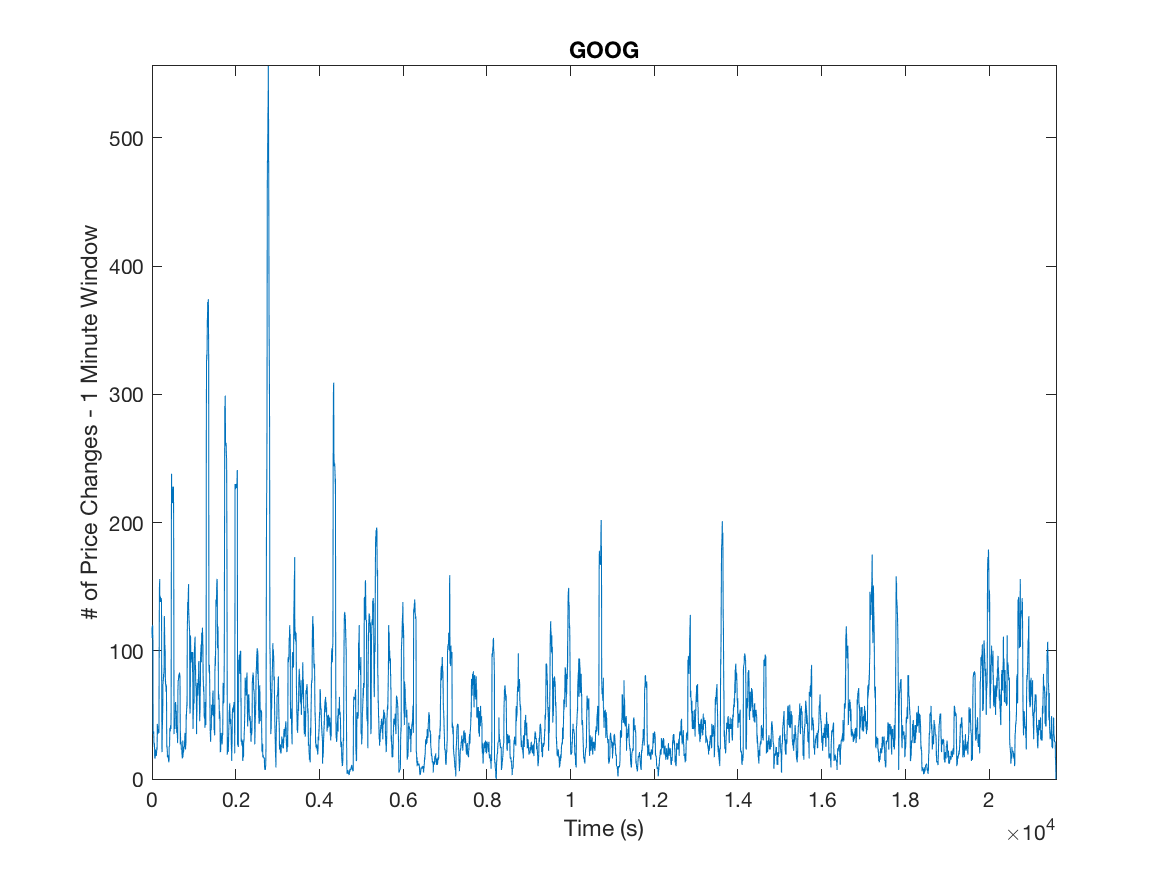}

\end{subfigure}
\begin{subfigure}[t]{0.49\textwidth}
\captionsetup{labelformat=empty}

\caption{\textbf{INTC}}
\includegraphics[width=\textwidth, trim = 0 0 0 30, clip]{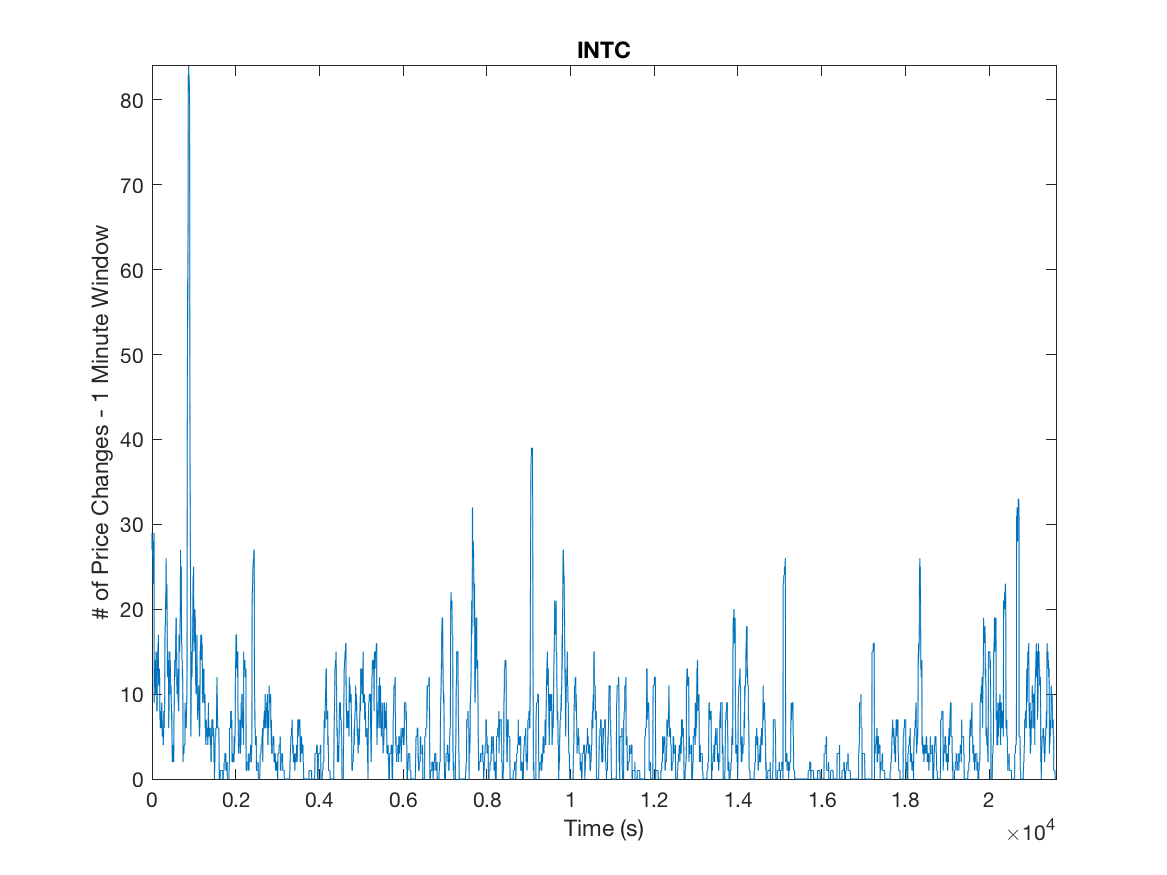}
\end{subfigure}

\vspace{3mm}

\begin{subfigure}[t]{0.49\textwidth}
\captionsetup{labelformat=empty}

\caption{MSFT}
\includegraphics[width=\textwidth, trim = 0 0 0 30, clip]{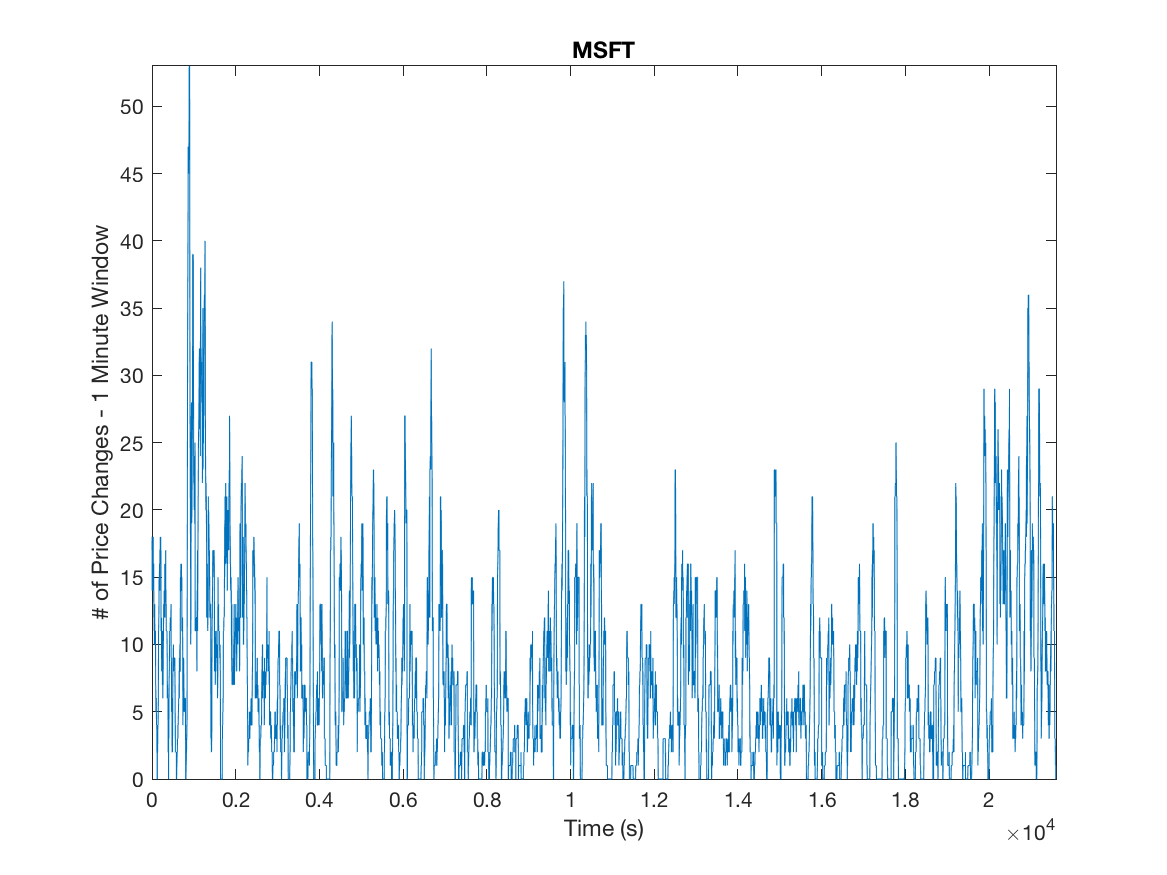}
\end{subfigure}

\caption{\label{fig:clustering} Each plot shows the number of arrivals for a moving one minute window. From this we can conclude that there is a significant amount of clustering in the arrival of mid-price changes, motivating the Hawkes model of our arrival process.}
\end{figure}

The high number of daily price changes motivates the idea that we can use asymptotic analysis in order to approximate long-run volatility using order flow by finding the diffusion limit of the price process. Because we do not want to include opening and closing auctions we omit the first and last fifteen minutes of our data. We motivate the arrival process by analyzing the inter-arrival times and clustering to a ensure the arrival process is not Poisson and exhibits the characteristics of a Hawkes Process, this is illustrated in Figures \ref{fig:poisson} and \ref{fig:clustering}.

Furthermore, the relationship between our diffusion coefficient and arrival process is limited to the expected number of arrivals on a unit interval. This implies that results for a simple exponential model can be easily generalized to a non-linear one. This makes it possible to work with a simplified model for our Hawkes process which is still rich enough to capture our observations. Keeping this in mind, we restrict ourselves to an exponential kernel and estimate parameters using a MLE \cite{35}. We provide these estimates in Table \ref{table:params} and compare the empirical expected number of arrivals and compare with the MLE estimate in Table \ref{table:unitinterval}

\begin{table}[htpb]

\centering
\begin{tabular}{c|c|c|c}
     &  $\lambda$ & $\alpha$ & $\beta$\\
\hline
AAPL & 1.4683 & 1045.2676 & 2556.1844\\
AMZN & 0.6443 & 653.7524 & 1556.1702\\
GOOG & 0.4985 & 865.8553 & 1980.4409\\
MSFT & 0.0659 &479.3482 & 908.0032\\
INTC & 0.0471 & 399.6389 & 760.4991
\end{tabular}

\caption{\label{table:params} Each parameter was estimated using a particle swarm optimization method in an attempt to globally optimize the negative log-likelihood function. The values for $\lambda$, $\alpha$ and $\beta$ for each data set are as provided.}
\end{table}

Obviously the MLE method accurately estimates the expected number of arrivals on the unit interval. This means we can confidently say for our data that our parameters will work reasonably with our models. We provide the estimated parameters in Table \ref{table:params}

\begin{table}[htpb]

\centering
\small

\begin{tabular}{ c | c | c }

 & Emp. $E[N([0,1])]$ & MLE\\
\hline
AAPL & 2.4840 & 2.4841\\

AMZN & 1.1110 & 1.1110\\

GOOG & 0.8857 & 0.8857\\

MSFT & 0.1395 & 0.1396\\

INTC & 0.0991 & 0.0992\\

\end{tabular}
\caption{\label{table:unitinterval} Expected number of arrivals on a unit interval using estimated parameters from an MLE method is compared against the empirical arrivals.}
\end{table}

\subsection{CHPDO}

We first consider a compound Hawkes process with dependent orders defined in Definition \ref{GCHPDO}, namely
\begin{equation}
    S_t = S_0 + \sum_{k=1}^{N(t)}X_k
\end{equation}
where $X_k\in\{+\delta, -\delta\}$ is a continuous-time two state Markov chain, $\delta$ is of fixed size and $N(t)$ is the number of mid-price movements up to time $t$ described by a Hawkes process.

We have opted to study the mid-price changes of our model. Thus $S_t$ can be computed by averaging the best bid and ask price. Noting that the price is recorded in cents the smallest possible jump in the mid-price is a half a cent which we will use as $\delta$. Furthermore, in order to estimate the transition matrix for the Markov chain $X_k$ we count the absolute frequencies of upward and downward price movements and from this calculate the relative frequencies giving an estimate for $p_{uu}$ and $p_{dd}$ which represent the conditional probabilities of an up/down movement given an up/down movement. Later we will consider several different sizes of mid-price movements and will work with the convention that each movement will be assigned a state based off of its ordering in the reals. In this case, $-\delta$ will be state one, and $\delta$ with be state two. This results in the transition matrix $P$ given below.

\[P =
\begin{bmatrix}
p_{dd} & 1-p_{dd}\\
1-p_{uu} & p_{uu}
\end{bmatrix}
\]

After determining our parameters and transition probabilities we calculate $a^*$ and $\sigma$ in Table \ref{table:chpdo_coeff} together with $p_{uu}$ and $p_{dd}$.

\begin{table}[htpb]

\centering
\small

\begin{tabular}{ c | c | c | c | c }

 & $p_{dd}$ & $p_{uu}$ & $\sigma$ & $a^*$\\
\hline
AAPL & 0.4956 & 0.4933 & 0.0049 & -1.1463e-5\\

AMZN & 0.4635 & 0.4576 & 0.0046 & -2.7373e-5\\

GOOG & 0.4769 & 0.4461 & 0.0046 & -1.4301e-4\\

MSFT & 0.6269 & 0.5827 & 0.0062 & -2.7956e-4\\

INTC & 0.6106 & 0.5588 & 0.0059 & -3.1185e-4\\

\end{tabular}

\caption{ \label{table:chpdo_coeff} Provided above are the values for $s^*$, $\sigma$ as well as the probabilities of an upward/downward movement given an upward/downward movement for each of the 5 stocks in question.}
\end{table}

\begin{figure}[htbp]

\centering
\begin{subfigure}[t]{0.49\textwidth}
\captionsetup{labelformat=empty}

\caption{\textbf{AAPL}}
\includegraphics[width=\textwidth, trim = 0 0 0 30, clip]{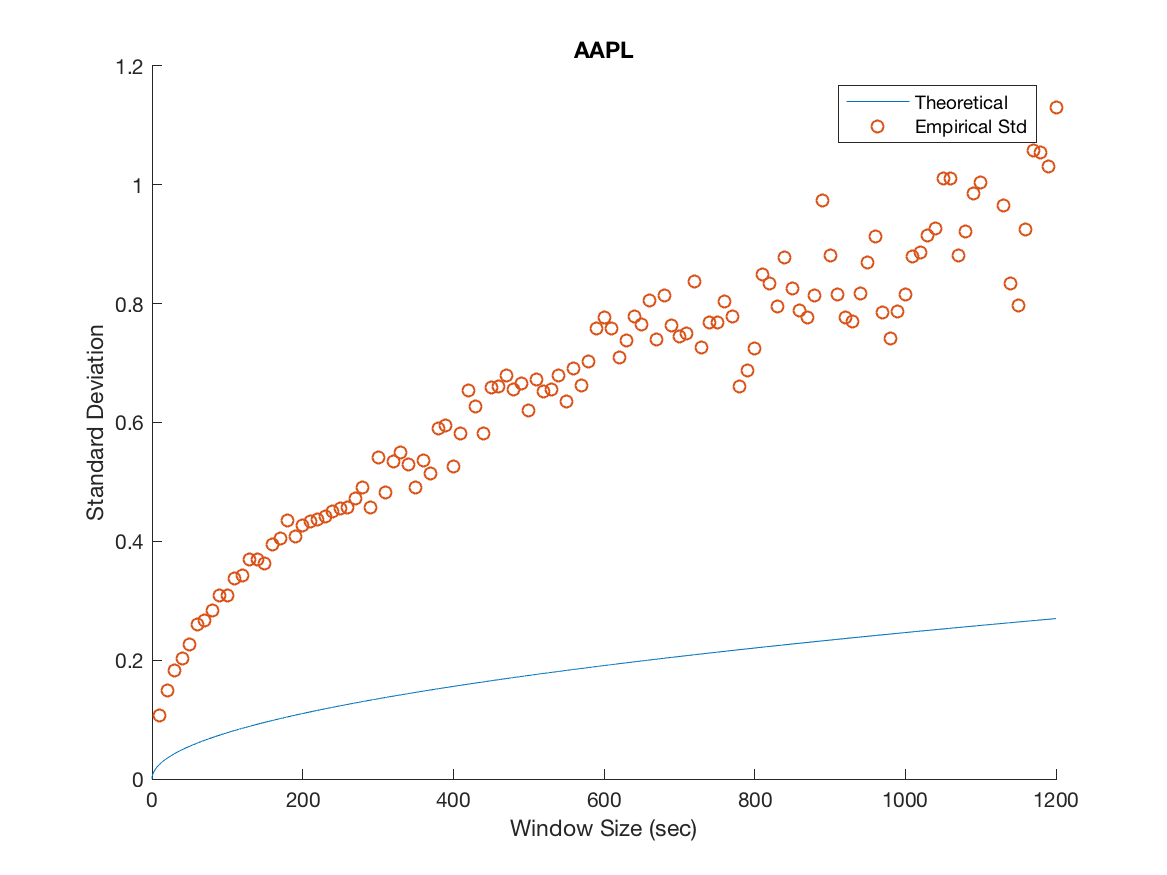}

\end{subfigure}
\begin{subfigure}[t]{0.49\textwidth}
\captionsetup{labelformat=empty}

\caption{\textbf{AMZN}}
\includegraphics[width=\textwidth, trim = 0 0 0 30, clip]{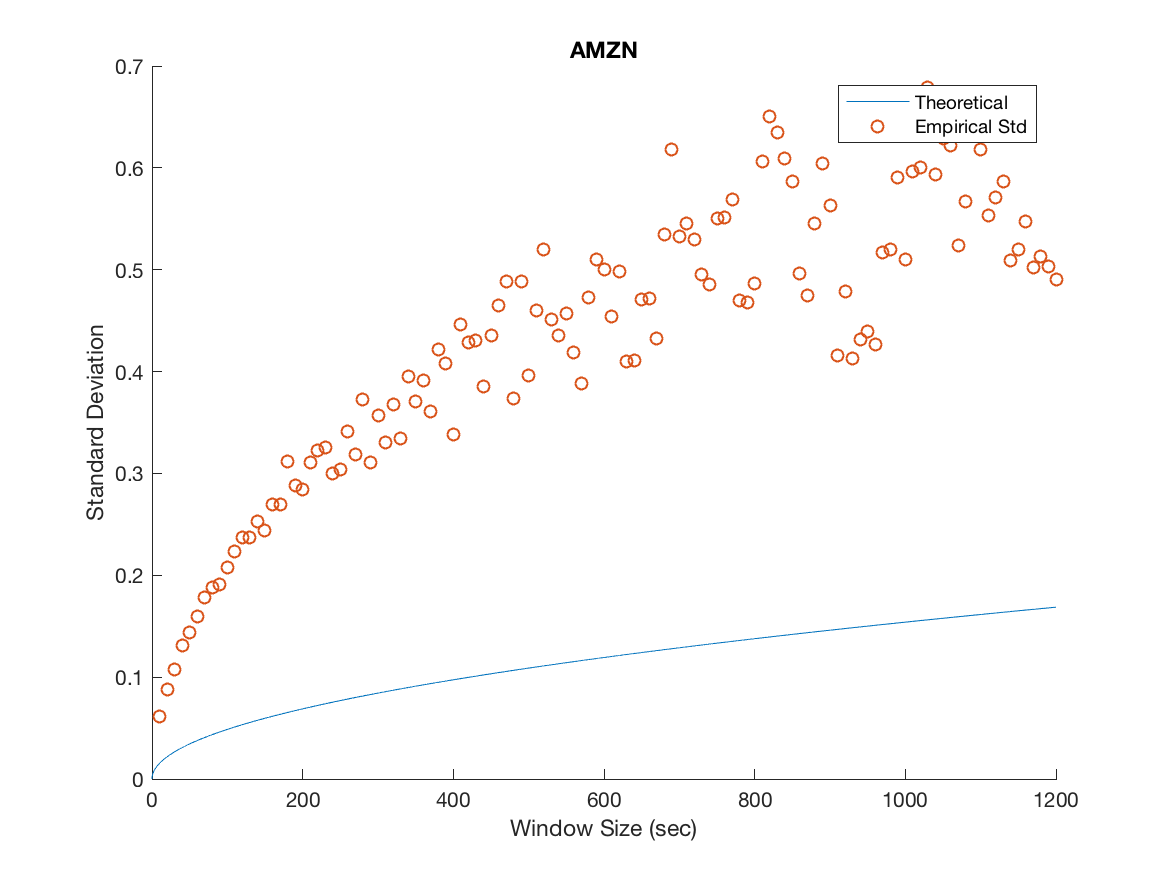}
\end{subfigure}

\vspace{3mm}

\begin{subfigure}[t]{0.49\textwidth}
\captionsetup{labelformat=empty}

\caption{\textbf{GOOG}}
\includegraphics[width=\textwidth, trim = 0 0 0 30, clip]{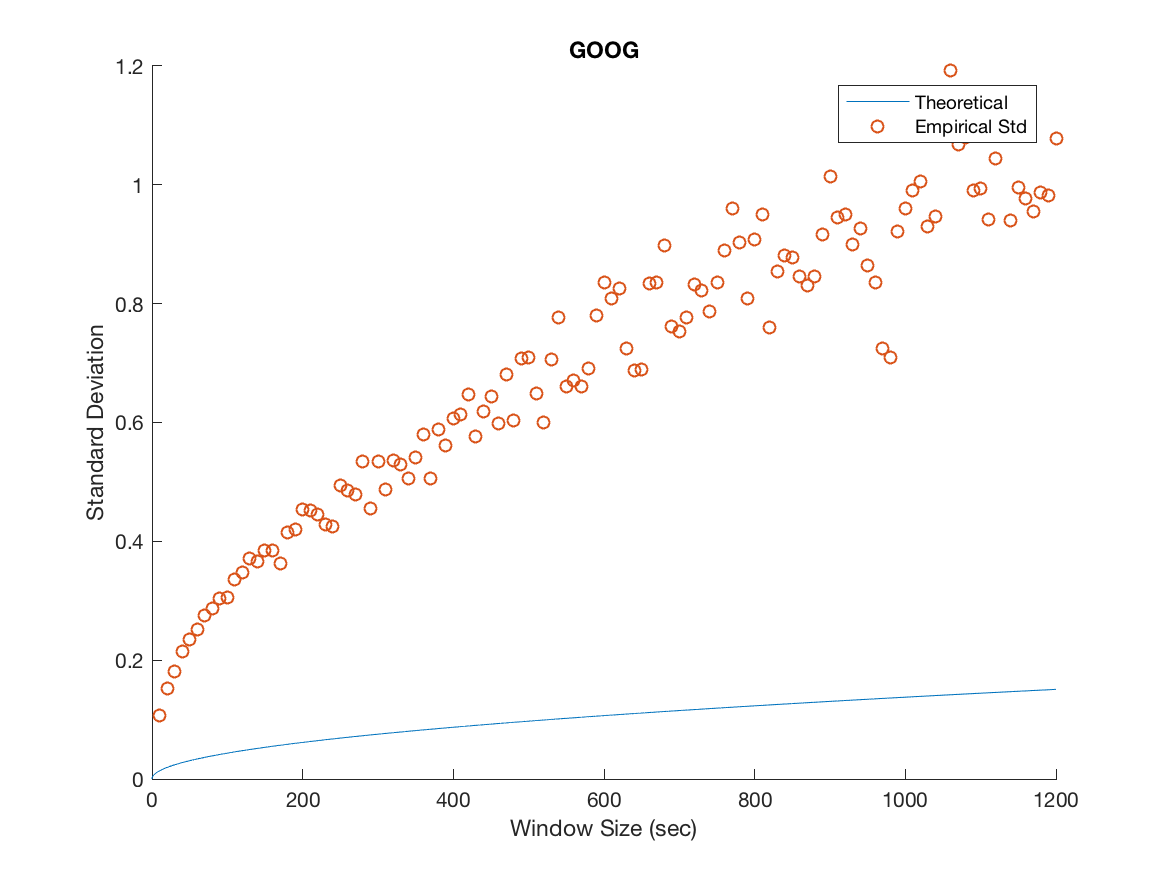}

\end{subfigure}
\begin{subfigure}[t]{0.49\textwidth}
\captionsetup{labelformat=empty}

\caption{\textbf{INTC}}
\includegraphics[width=\textwidth, trim = 0 0 0 30, clip]{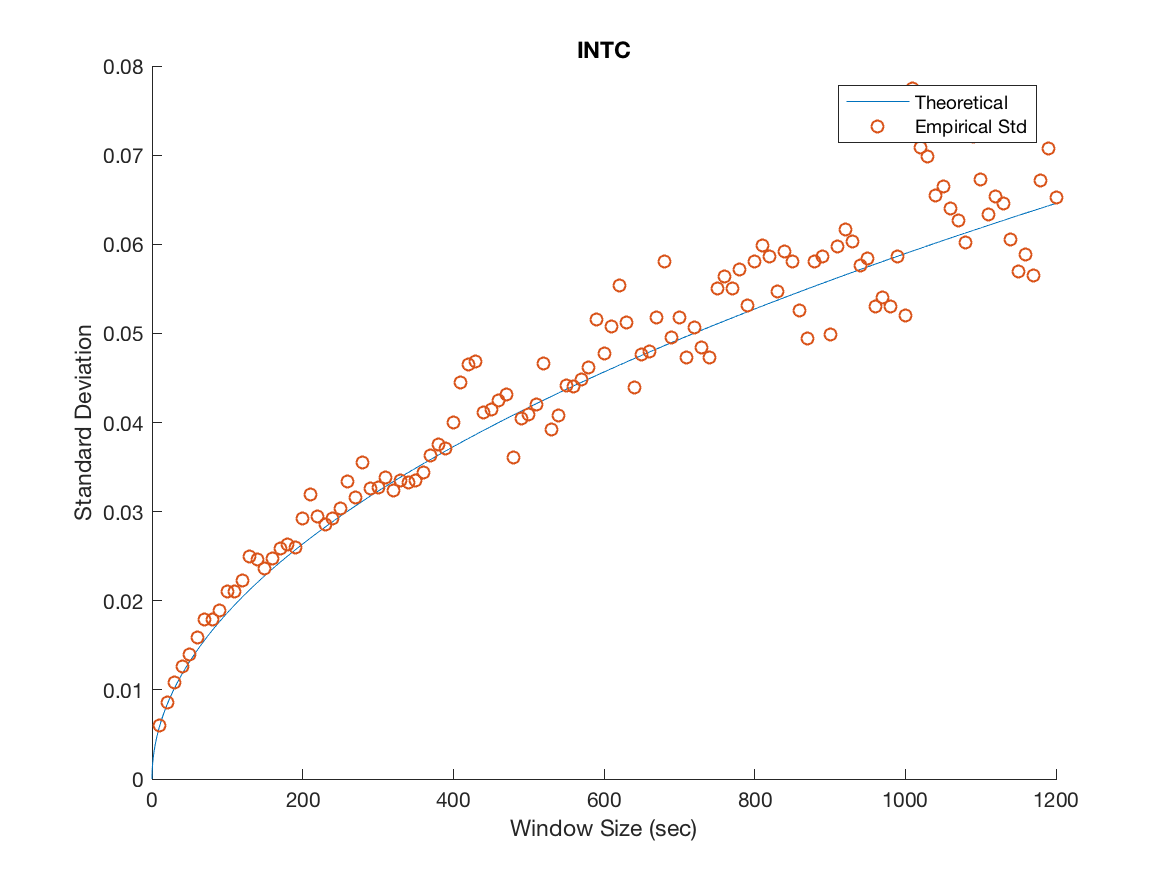}
\end{subfigure}

\vspace{3mm}

\begin{subfigure}[t]{0.49\textwidth}
\captionsetup{labelformat=empty}

\caption{MSFT}
\includegraphics[width=\textwidth, trim = 0 0 0 30, clip]{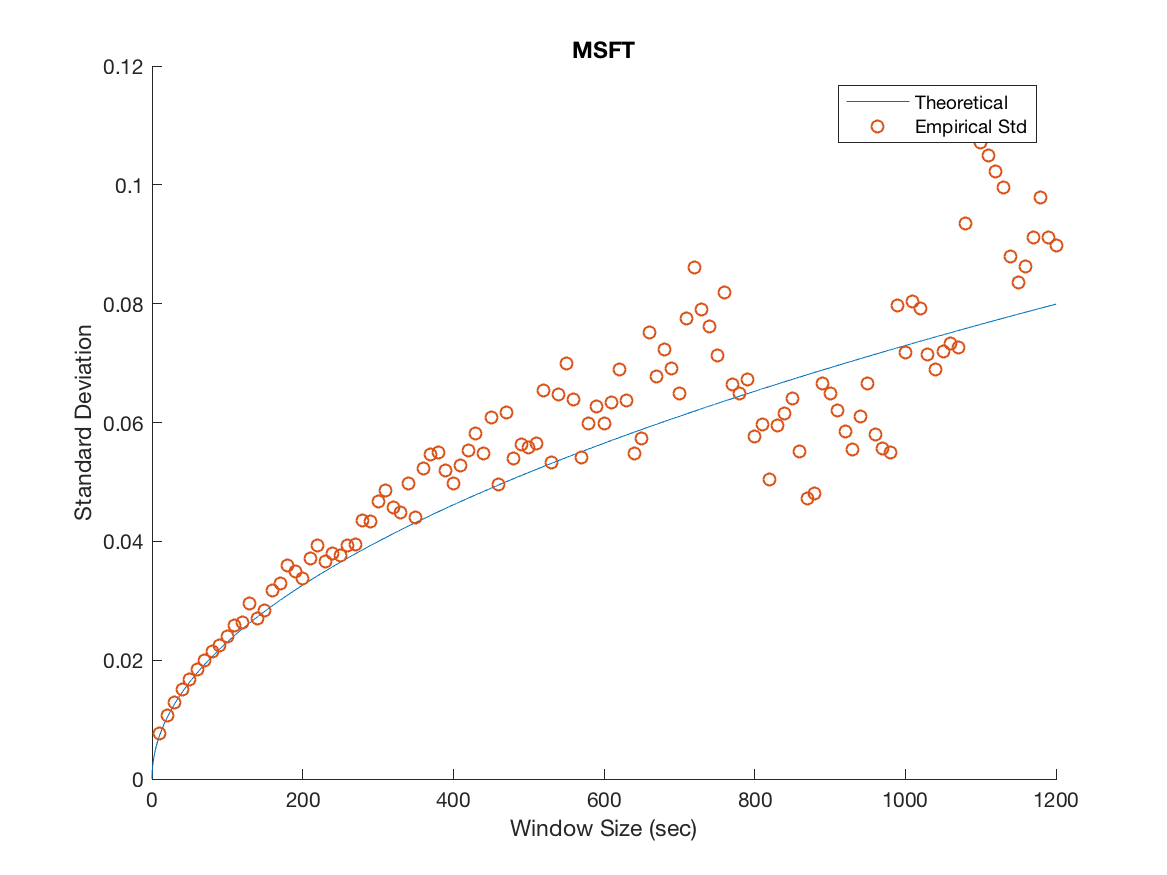}
\end{subfigure}

\caption{\label{fig:chpdofit} Each figure compares the empirical standard deviation for a fixed window size to the theoretical standard deviation. We have plotted an empirical standard deviation for all $n$ from 10 seconds to 20 minutes in step sizes of 10 seconds. Each empirical standard deviation corresponds to a single point in the scatter plot and the plotted curve corresponds to the predicted theoretical value.}

\end{figure}

Provided these values we can test our claim that our model accurately describes the mid-price process satisfies we will use the diffusion limit proved earier, namely
\begin{equation}
    \frac{S_{nt} - N(nt)a^*}{\sqrt{n}}\xrightarrow{n\to\infty} \sigma\sqrt{\frac{\lambda}{1-\alpha/\beta}}W(t).
\end{equation}
If the data satisfies our proposed model then when considering large windows of time (5min, 10min, 20min), then when we would expect to see the empirical and theoretical standard deviations to follow each other closely. To test this we compare the equivalent process, constructed by multiplying the LHS and RHS by $\sqrt{n}$. Then cutting our data into disjoint windows of size $n$, specifically $[in, (i+1)n]$ with $t = 1$ and by setting the left bound as our starting time we can calculate $S_{nt} - N(nt)a^*$ for each individual window and give a generalized formula for this below.
\begin{equation}\label{eq:S_star}
    S^*_{i} = S_{(i+1)nt} - S_{int} - (N((i+1)nt) - N(int))a^*
\end{equation}
 This gives a collection of values $\{S^*_i\}$ over which we compute the standard deviation. If our model is accurate would would expect that 
 \begin{equation}
    \text{std}\{S^*_i\} \approx \sqrt{nt}\sigma\sqrt{\frac{\lambda}{1-\alpha/\beta}},\ \text{where } t = 1.
 \end{equation}
 We plot the empirical standard deviation against the theoretical one for various window sizes starting at 10 seconds and increasing in steps of 10 seconds until we reach 20 minutes, this is illustrated in Figure \ref{fig:chpdofit}.

Several important remarks should be made at this point. It is clear that while the model accurately predicts the overall trend for MSFT and INTC, we severely underestimate the variability in the mid-price process for APPL, AMZN and GOOG. Furthermore, as the window size increases the overall spread in the data increases. We attribute this to the decreasing sample size imposed on us as we increase the window size. For example when we consider a 20 minute window, we can only construct 27 disjoint windows in the 9 hour trading day forcing us to deal with the problem of predicting a 'population' standard deviation from a increasingly small sample. We remedy this by using a variance stabilizing transformation in later sections. Specifically, a popular method for a Poisson process is to take the square root of our empirical and theoretical standard deviations. This makes it possible to qualitatively view the overall trend in the data, gaining a clearer idea of goodness of fit from there.

\subsection{GCHP2SDO}

\begin{figure}[htbp]

\centering
\begin{subfigure}[t]{0.49\textwidth}
\captionsetup{labelformat=empty}

\caption{\textbf{AAPL}}
\includegraphics[width=\textwidth, trim = 0 0 0 30, clip]{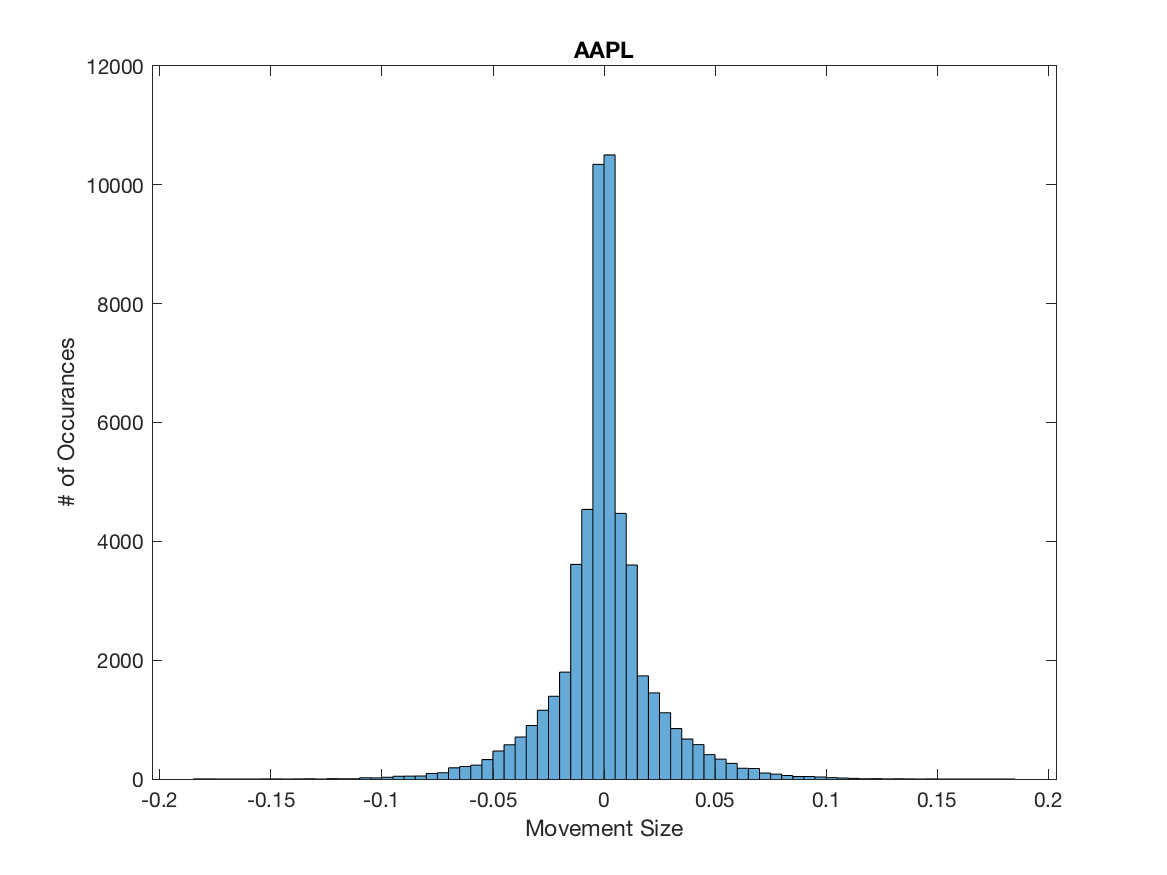}

\end{subfigure}
\begin{subfigure}[t]{0.49\textwidth}
\captionsetup{labelformat=empty}

\caption{\textbf{AMZN}}
\includegraphics[width=\textwidth, trim = 0 0 0 30, clip]{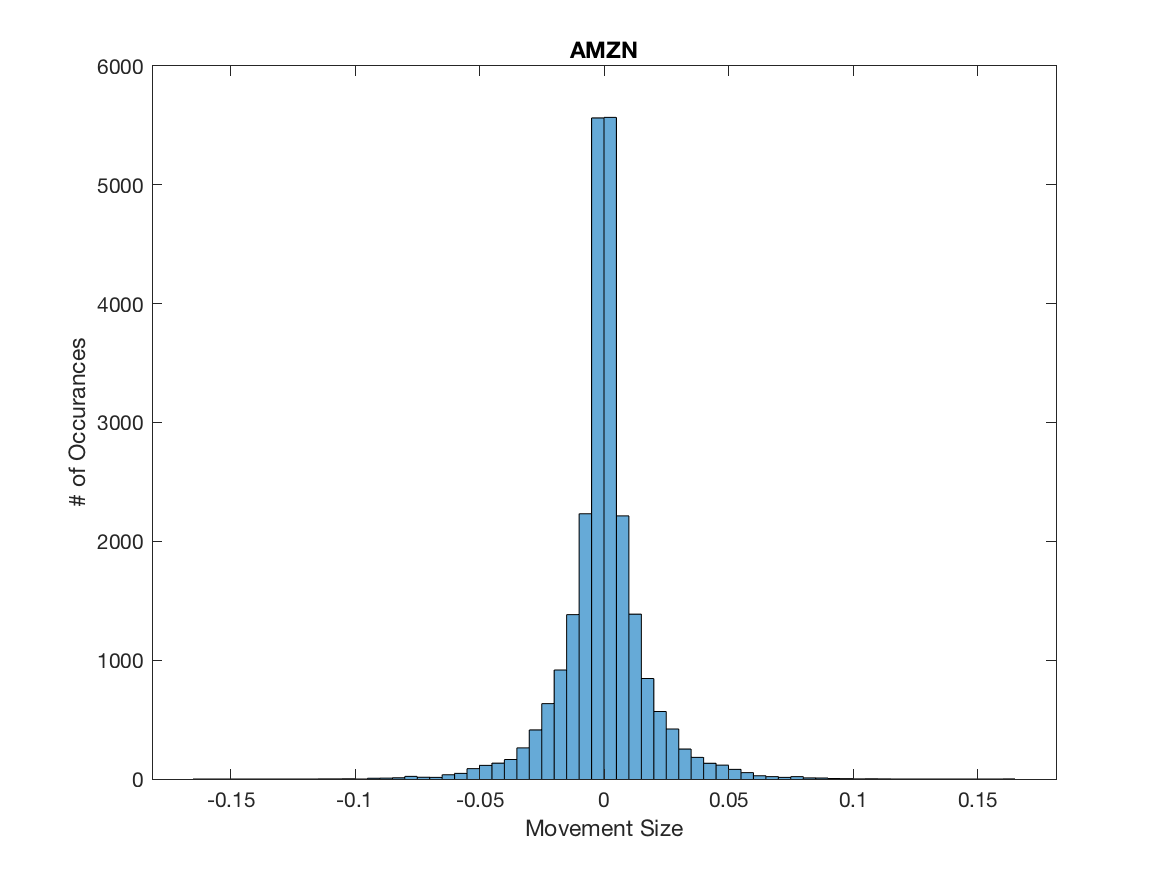}

\end{subfigure}

\begin{subfigure}[t]{0.49\textwidth}
\captionsetup{labelformat=empty}

\caption{\textbf{GOOG}}
\includegraphics[width=\textwidth, trim = 0 0 0 30, clip]{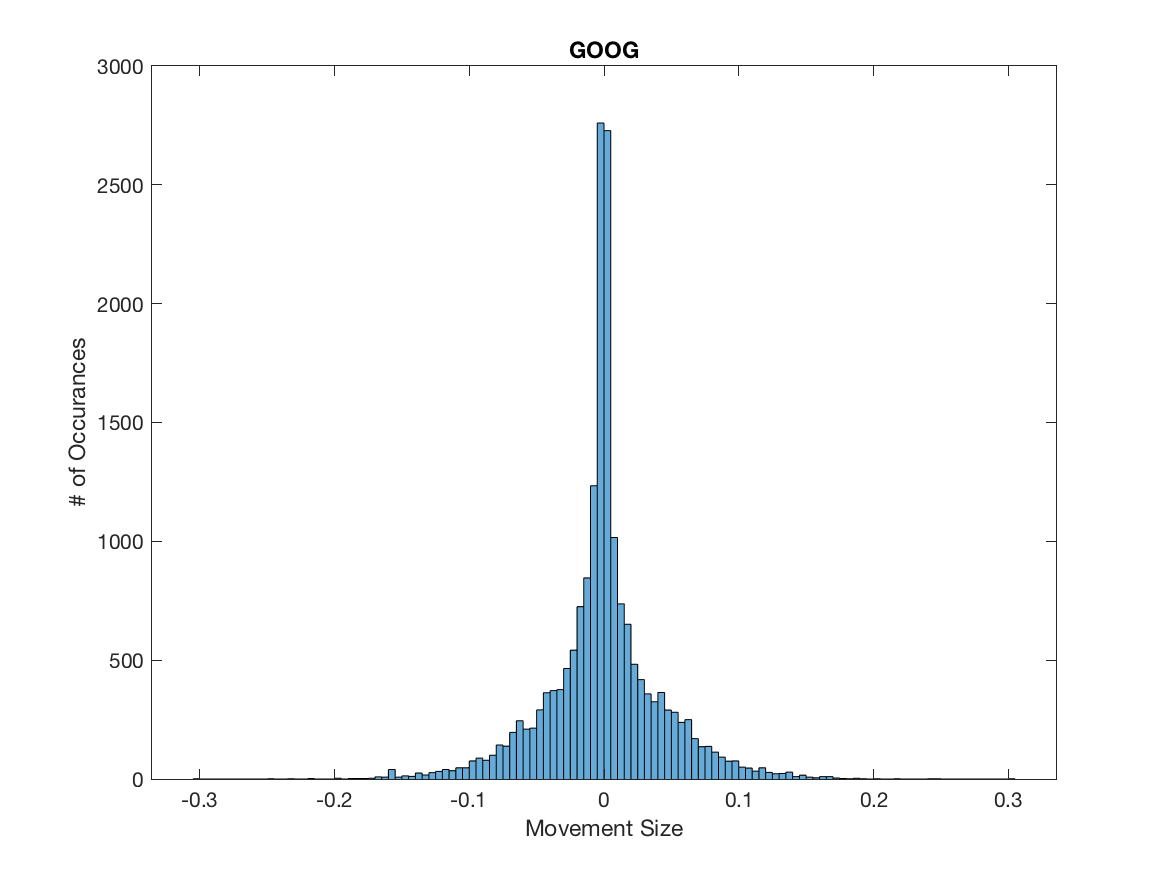}

\end{subfigure}

\caption{\label{fig:tickhist} We can see clearly that the change in the mid-price is often larger than a half tick. These mid-price changes make up a significant portion of the actual data, contradicting the assumption needed for the CHPDO model that the mid-price changes occur on average at a half tick size.}
\end{figure}

As of now we have considered a fixed delta related to the trading tick size. However, if we consider the mid-price changes for APPL, AMZN and GOOG the assumption of a fixed tick size is violated. In fact, we observe that approximately approximately 61\%, 53\% and 71\% of all mid-price changes are larger than half a tick size, which is opposed to what we observe for MSFT and INTC where all mid-price changes occur at the half tick size, we illustrate this for AAPL, AMZN and GOOG in Figure \ref{fig:tickhist}.

It is clear in Figure \ref{fig:tickhist} additional considerations need to be made. A simple way to include the variability in mid-price movements in our model is to introduce $a(X_i)$ as described in Definition \ref{GCHP2SDO}. It is of course necessary to determine the values of $a( \cdot )$ for each state of our Markov chain. A naive method is to take the mean of the downward and upward mid-price movements and assign them to $a(1)$ and $a(2)$ respectively. We provide these values in Table \ref{table:2sdoticks}.

\begin{table}[htpb]

\centering
\begin{tabular}{c|c|c}
     &  $a(1)$ & $a(2)$\\
\hline
AAPL & -0.0172 & 0.0170\\
\hline
AMZN & -0.0134 & 0.0133\\
\hline
GOOG & -0.0302 & 0.0308\\
\end{tabular}

\caption{\label{table:2sdoticks} $a(i)$ is the average of the upward or downward mid-price movements. Following our previous convention, the first state will be associated with the mean of all downward mid-price movements and the second state will be associated with the mean of all upward mid-price movements.}
\end{table}

\begin{figure}[htbp]

\centering

\begin{subfigure}[t]{0.49\textwidth}
\captionsetup{labelformat=empty}

\caption{\textbf{AAPL}}
\includegraphics[width=\textwidth]{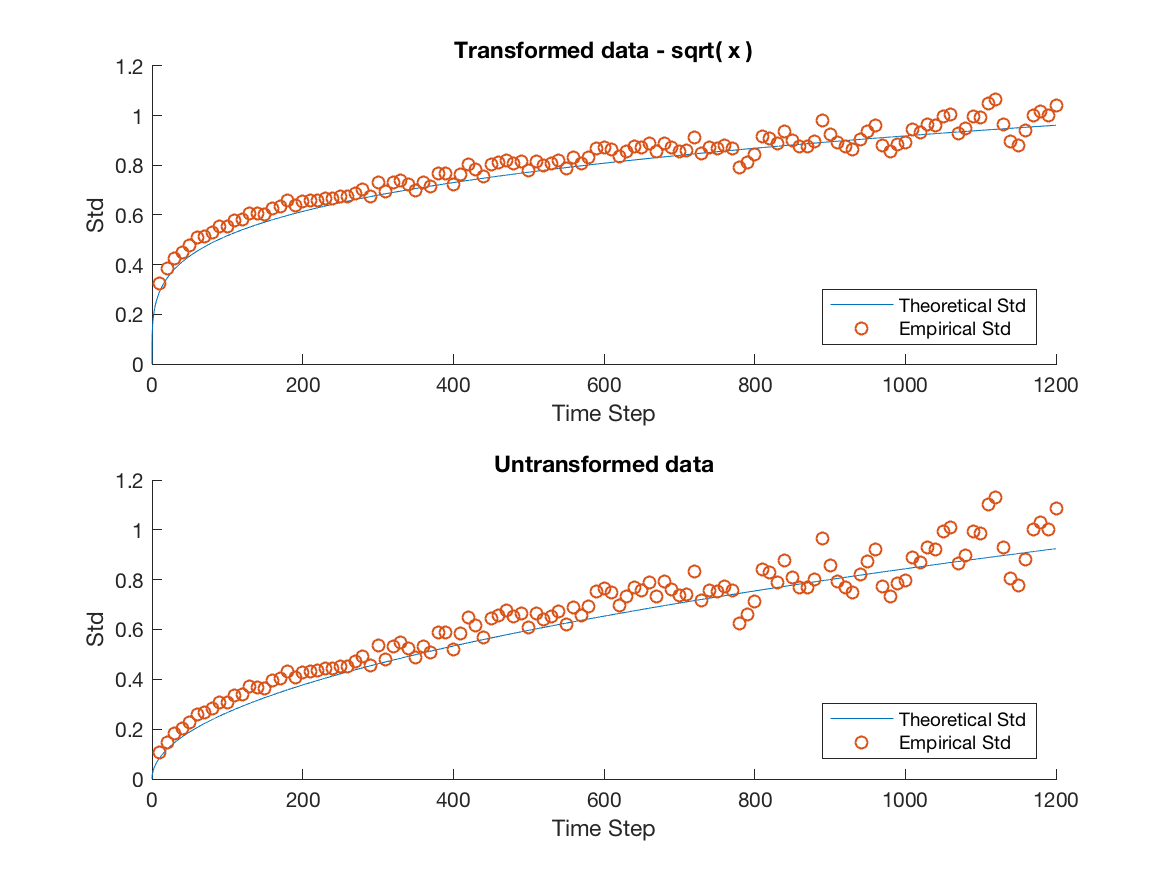}

\end{subfigure}
\begin{subfigure}[t]{0.49\textwidth}
\captionsetup{labelformat=empty}

\caption{\textbf{AMZN}}
\includegraphics[width=\textwidth]{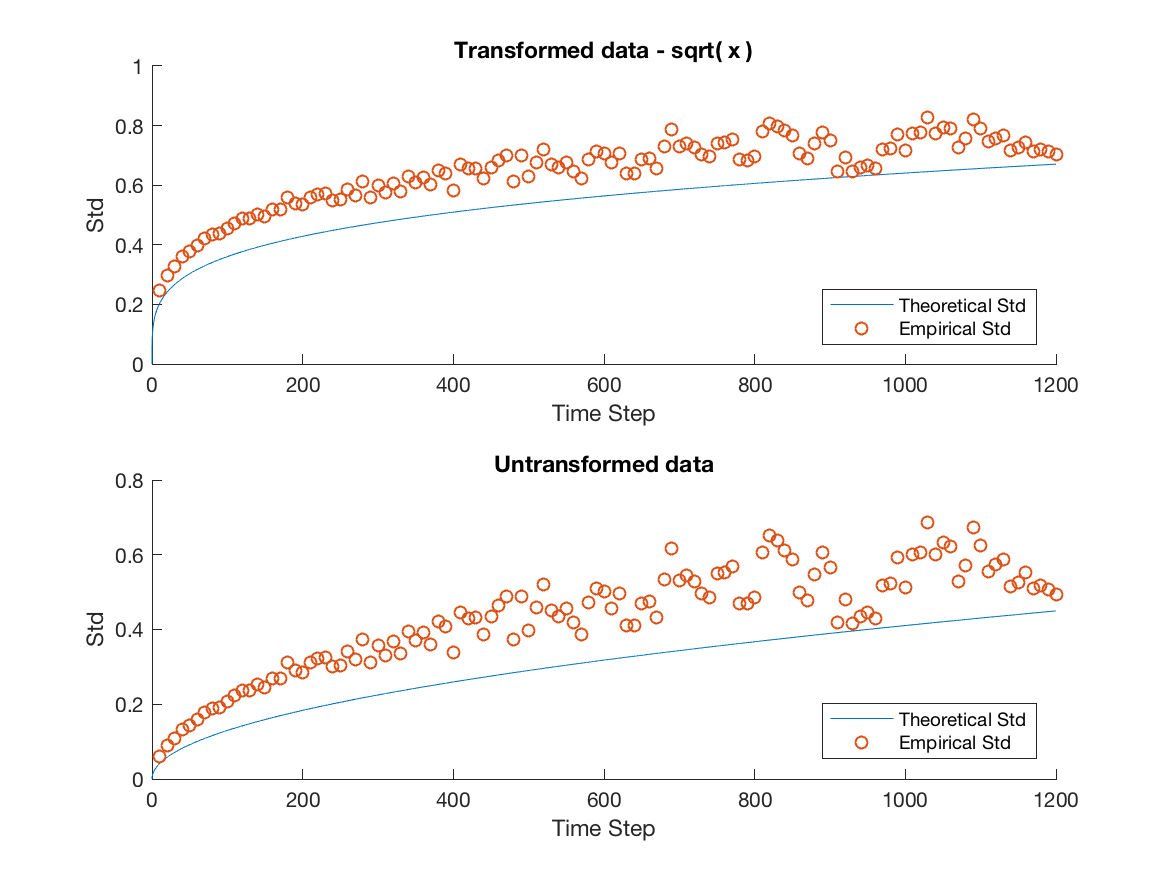}

\end{subfigure}

\begin{subfigure}[t]{0.49\textwidth}
\captionsetup{labelformat=empty}

\caption{\textbf{GOOG}}
\includegraphics[width=\textwidth]{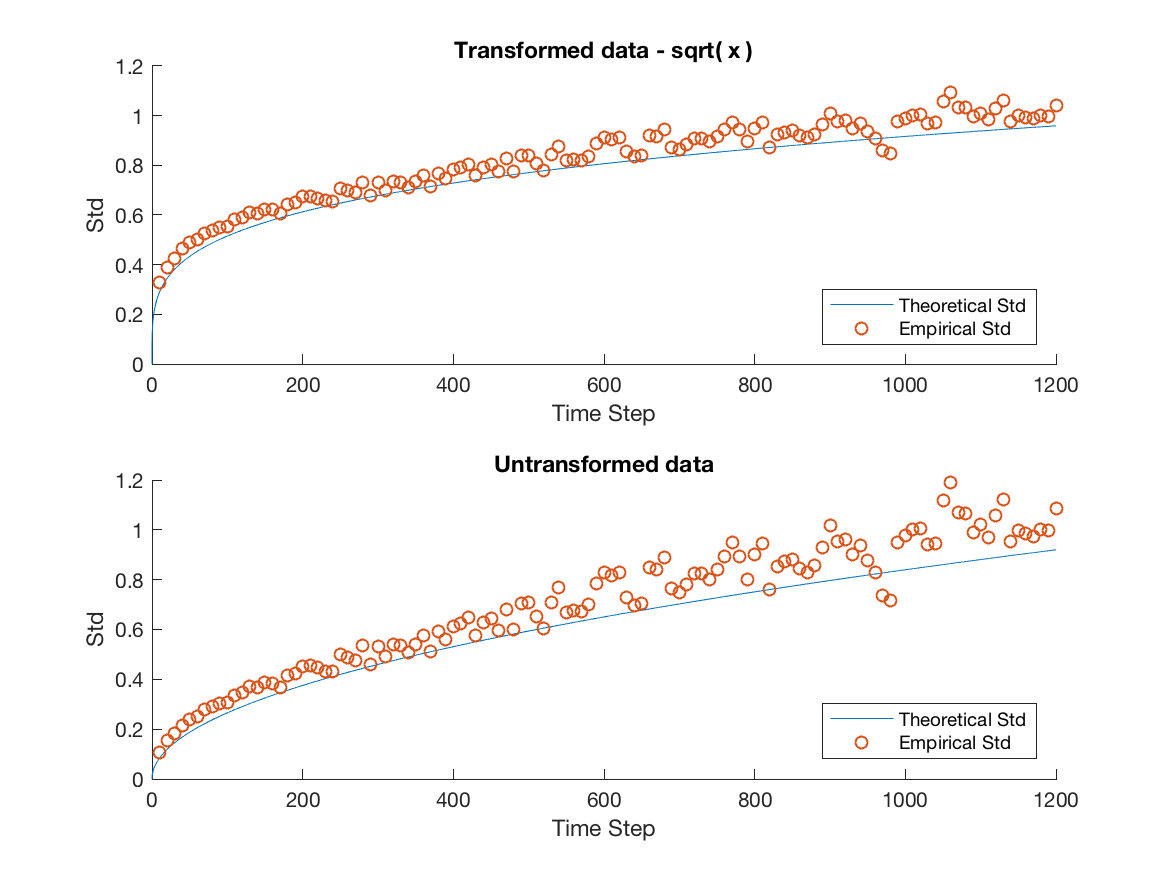}

\end{subfigure}

\caption{\label{fig:2SDOfit} A comparison of the empirical standard deviation for a fixed window size $n$ to the theoretical standard deviation for AAPL, AMZN and GOOG using the 2 state dependent order model. We have plotted the empirical standard deviation for all $n$ from 10 seconds to 20 minutes in step sizes of 10 seconds. Each empirical standard deviation corresponds to a single point in the scatter plot and the plotted curve corresponds to the predicted theoretical value. Visually there is a significant improvement for all stocks, although the theoretical standard deviation for AMZN is still underestimating the empirical variability.}

\end{figure}

In this step we have only endeavoured to better realize the actual price movements in our data. Therefore, when we observe a downward mid-price movement we continue to assign it to state one and similarly for an upward price movement we continue to assign it to state two. It follows that our transition matrix will remain the same. Then using these new state values we recalculate $a^*$ and $\sigma$, providing them in Table \ref{table:2SDO_coeff}. The effect of these changes is investigated in Figure \ref{fig:2SDOfit}. Note that in Figure \ref{fig:2SDOfit} we have used the variance stabilizing transformation discussed earlier in order to better visualize the overall trend in our data.

\begin{table}[htpb]

\centering
\small

\begin{tabular}{ c | c | c | c | c }
 & $p_{dd}$ & $p_{uu}$ & $\sigma$ & $a^*$\\
\hline
AAPL & 0.4956 & 0.4933 & 0.0169 & -1.5624e-4\\
AMZN & 0.4635 & 0.4576 & 0.0123 & -1.0475e-4\\
GOOG & 0.4769 & 0.4461 & 0.0282 & -5.5095e-4\\
\end{tabular}

\caption{\label{table:2SDO_coeff} We above the values for $s^*$, $\sigma$ as well as the probabilities of an upwards/downwards movement given an upwards/downwards movement for the 3 stocks of interest.}
\end{table}

Notice that there is a significant qualitative improvement in the fits for AAPL and GOOG in Figure \ref{fig:2SDOfit} but the variability in mid-price movements for AMZN is still clearly underestimated by our model. The unexplained variance may be captured by investigating an n-state Markov chain since the additional transition probabilities could explain the variability missing in the 2-state case.

\subsection{GCHPNSDO}

We recall the N-state model described in Definition \ref{GCHPnSDO}. The immediate question becomes how best to choose the state values. We modify the quantile based approach from \cite{45}. After calculating the mid-price changes we separate the data into upward and downward price movements. Then we calculate evenly distributed quantiles for both data sets. Depending on the data, several quantiles may be identical, we reject any duplicates. We thus obtain a list of bounds which we complete by adding the minimum observed value if necessary.

To determine the state values $a(X_i)$, we take the average of all mid-price changes located between two neighbouring boundary values. Furthermore, we assign a mid-price change to state $i$ if it is greater than or equal to the $(i-1)$th boundary and strictly less than the $i$th boundary. An exception is made for the largest upper bound where equality is permitted at both ends.

\begin{figure}
    \centering
    
    \caption*{\textbf{AMZN}}
    \includegraphics[width=\textwidth]{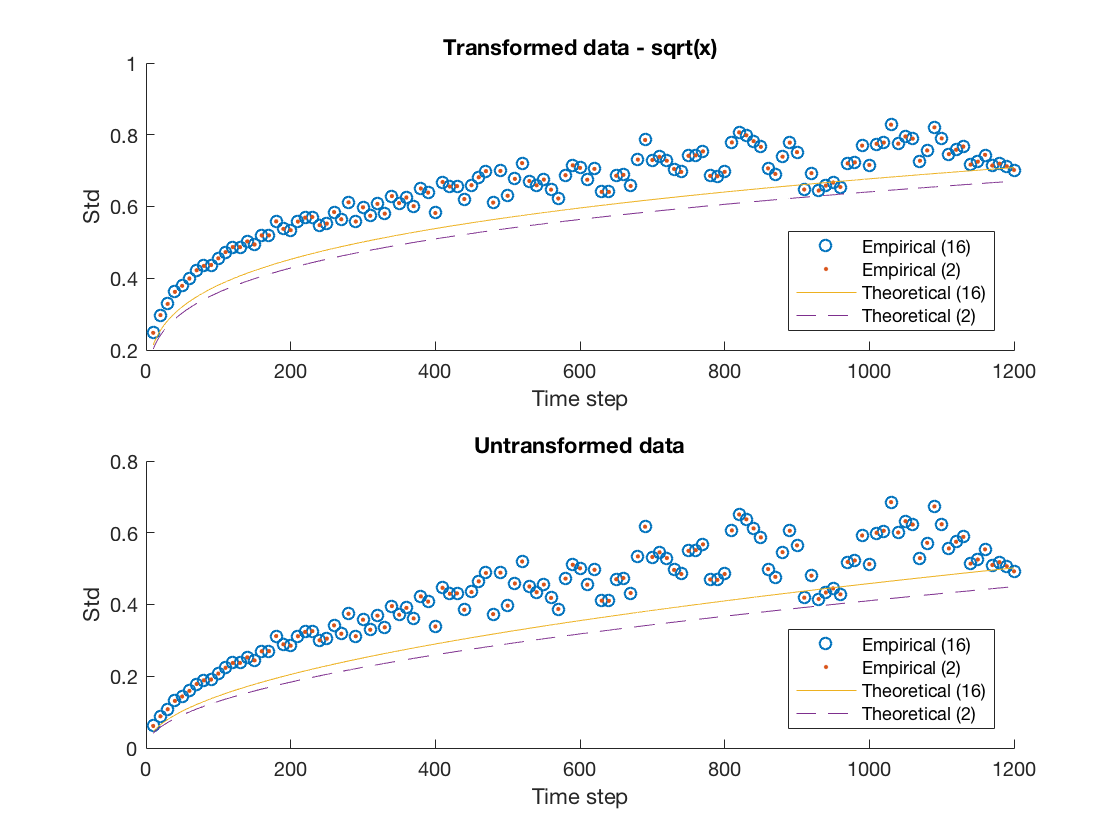}
    
    \caption{\label{fig:AMZN_data} We consider the N-state model for AMZN discussed previously in the paper. While there is a slight improvement against the original fit, the model still struggles to perfectly predict the variability in the mid-price changes of our data.}
    
\end{figure}

As we could not capture the full variability of mid-price changes for AMZN in the previous method we investigate for this case. Furthermore, for tractability we only consider 14 boundary values from which we obtain a 12-state Markov chain. Instead of providing the transition matrix, we provide the ergodic probabilities for the transition matrix and the associated states in Table \ref{table:AMZN_data}.
\begin{table}[htpb]

\centering
\small
\textbf{AMZN}

\vspace{\baselineskip}

\begin{tabular}{ c | c | c }
i & $\pi_i^*$ & $a(X_i)$\\
\hline
1 & 0.0275 & -0.0524\\
2 & 0.0281 & -0.0318\\
3 & 0.0264 & -0.0250\\
4 & 0.0382 & -0.0200\\
5 & 0.0576 & -0.0150\\
6 & 0.3249 & -0.0064\\
7 & 0.2321 & 0.0050\\
8 & 0.0923 & 0.0100\\
9 & 0.0578 & 0.0150\\
10 & 0.0353 & 0.0200\\
11 & 0.0412 & 0.0271\\
12 & 0.0387 & 0.0476
\end{tabular}

\caption{\label{table:AMZN_data} Above we have provide the state, associated ergodic probabilities and state values $a(i)$ for AMZN, given a 12 state Markov chain which was obtained from choosing a 16 quantile method.}
\end{table}
In order to compare the two state and N-state approaches we first take a qualitative approach and plot the two theoretical and empirical standard deviations against each other in Figure \ref{fig:AMZN_data}. When we compare the mean squared residuals, the 2-state model discussed before has mean squared error $0.0208$ while our 12-state Markov chain brings that down to $0.0125$. Considering an even larger Markov chain with 24-states we are only able to obtain a meager improvement to $0.0123$, suggesting that there is some underlying variance in the mid-price process not captured by our model. We investigate these more quantitative measures in the following Subsection. First comparing the models against a numerical best fit and then investigating the mean squared error to gain a better quantitative understanding of the overall improvement obtained from each model as we increase the number of quantiles.

\subsection{Quantitative Analysis}

While our model does visually appear to fit the expected variability in four of the five cases, it still fails to capture the complete dynamics of mid-price changes seen in our AMZN data. We investigate the mean square error of our models with a varying number of quantiles in Table \ref{table:meanres}, this gives a good indication as to whether the N-state model is a better fit for our data. If we look closely at AAPL and GOOG we see the N-state case can still improve our results from the two state case. For AAPL we constructed a 17 state Markov chain by taking 16 quantiles on the downward movements and 16 quantiles upward movements. This resulted in a mean squared error of $0.0036$ which is approximately a 28\% improvement to the two state case where the mean squared error was $0.0050$. Even more extreme, using a 25 state Markov chain for GOOG which was constructed similarly, we observed a mean squared error of $0.0046$ which is a 60\% improvement from the two state case with a mean squared error of $0.0115$. We conclude with Table \ref{table:meanres} which provides the mean residuals for AAPL, AMZN and GOOG with several of Markov chains constructed from various numbers of quantiles. We also include mean residuals for INTC and MSFT for comparison.

\begin{table}[htpb]
    \centering
    \begin{tabular}{c|c|c|c|c|c}
     & CHPDO & 2 & 8 & 16 & 32\\
    \hline
    AAPL & 0.2679 & 0.0050 & 0.0036 & 0.0036 & 0.0036\\
    AMZN & 0.1122 & 0.0208 & 0.0131 & 0.0124 & 0.0123\\
    GOOG & 0.4036 & 0.0115 & 0.0048 & 0.0045 & 0.0047\\
    INTC & 1.7917e-5 & 1.7917e-5 & 1.7917e-5 & 1.7917e-5 & 1.7917e-5\\
    MSFT & 1.0586e-4 & 1.0586e-4 & 1.0586e-4 & 1.0586e-4 & 1.0586e-4
    \end{tabular}
    \caption{\label{table:meanres} We list the mean residuals for several Markov chains with varying numbers of states. These were generated using our modified quantile approach choosing to start with 2, 8, 16 or 32 quantiles. We see that in general the mean residual decreases to some lower limit where we can no longer perform any better. Recall that the only observed mid-price changes for INTC and MSFT were of a half tick size, any increase in the number of quantiles will result in the same performance.}
    
\end{table}

\begin{figure}[htbp]

\centering

\begin{subfigure}[t]{0.49\textwidth}
\captionsetup{labelformat=empty}

\caption{\textbf{AAPL}}
\includegraphics[width=\textwidth, trim = 0 0 0 30, clip]{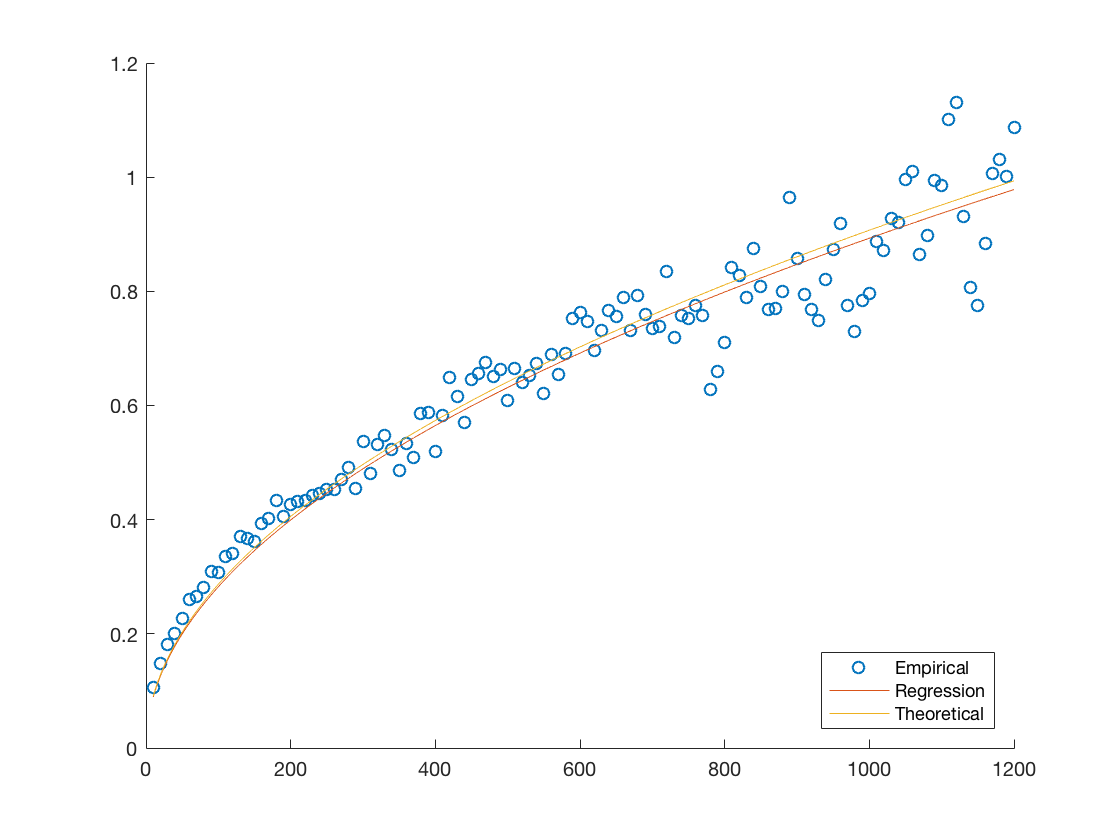}

\end{subfigure}
\begin{subfigure}[t]{0.49\textwidth}
\captionsetup{labelformat=empty}

\caption{\textbf{AMZN}}
\includegraphics[width=\textwidth, trim = 0 0 0 30, clip]{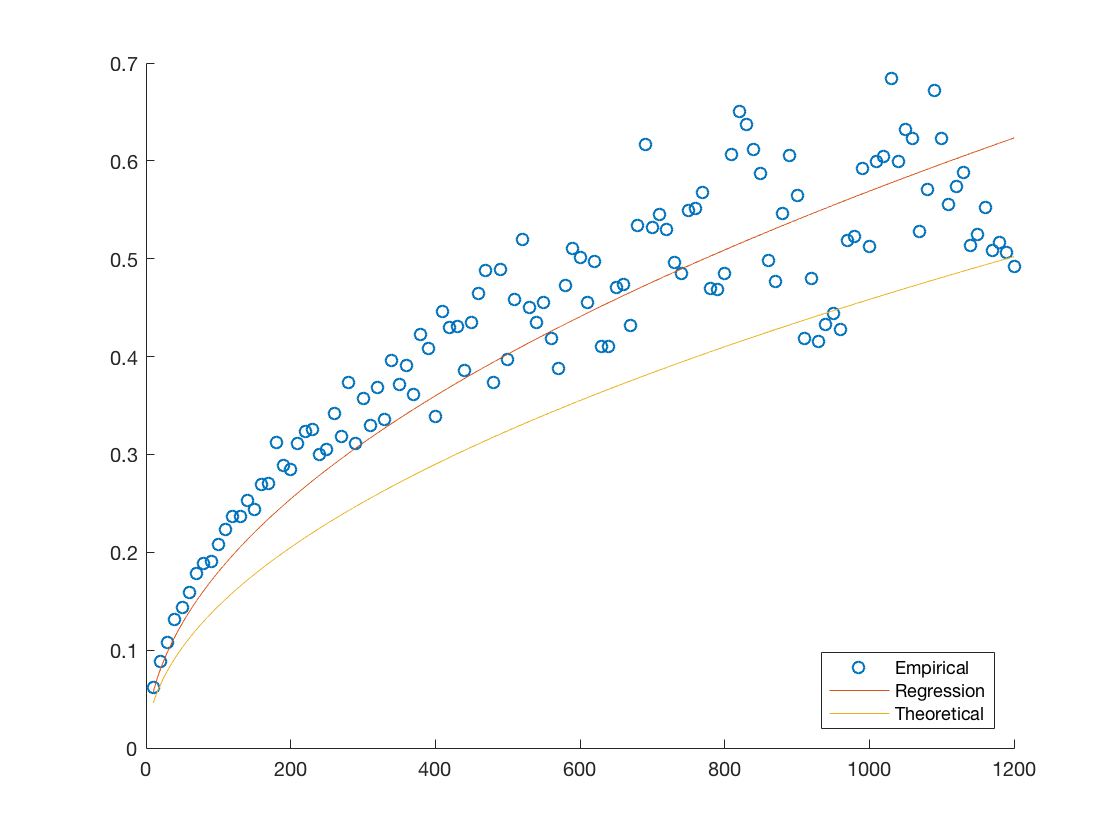}
\end{subfigure}

\vspace{3mm}

\begin{subfigure}[t]{0.49\textwidth}
\captionsetup{labelformat=empty}

\caption{\textbf{GOOG}}
\includegraphics[width=\textwidth, trim = 0 0 0 30, clip]{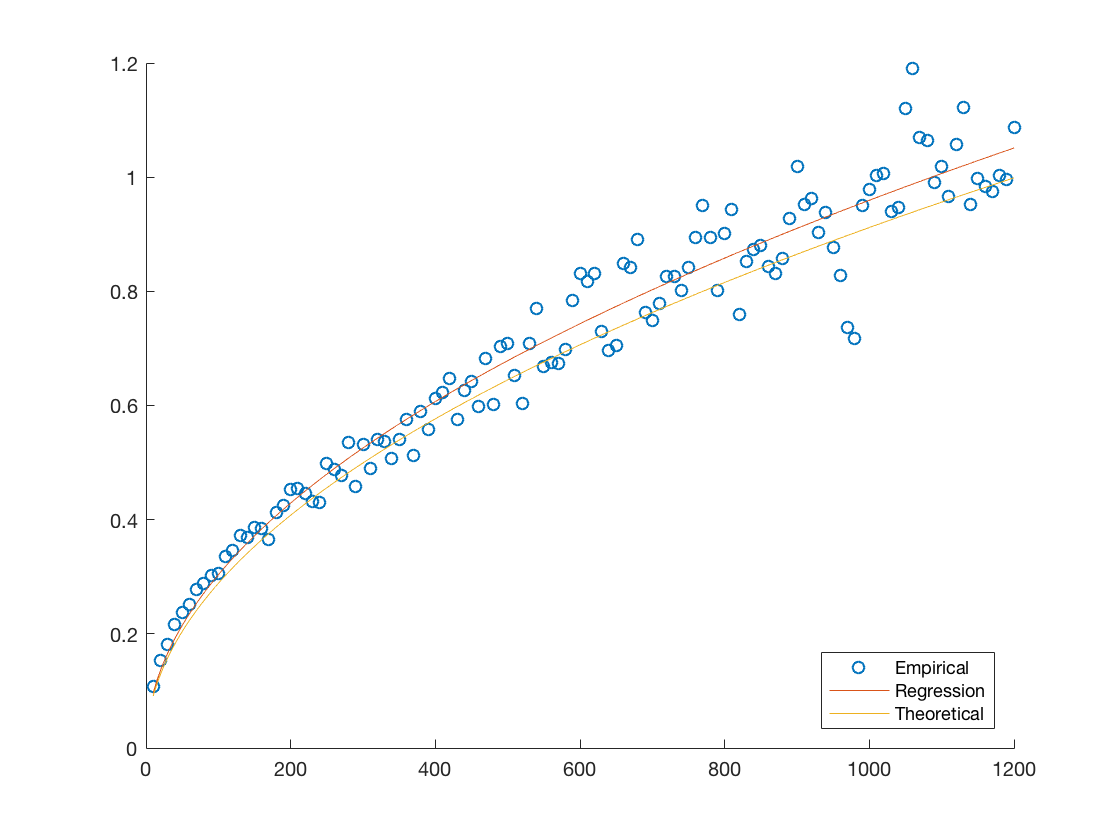}

\end{subfigure}
\begin{subfigure}[t]{0.49\textwidth}
\captionsetup{labelformat=empty}

\caption{\textbf{INTC}}
\includegraphics[width=\textwidth, trim = 0 0 0 30, clip]{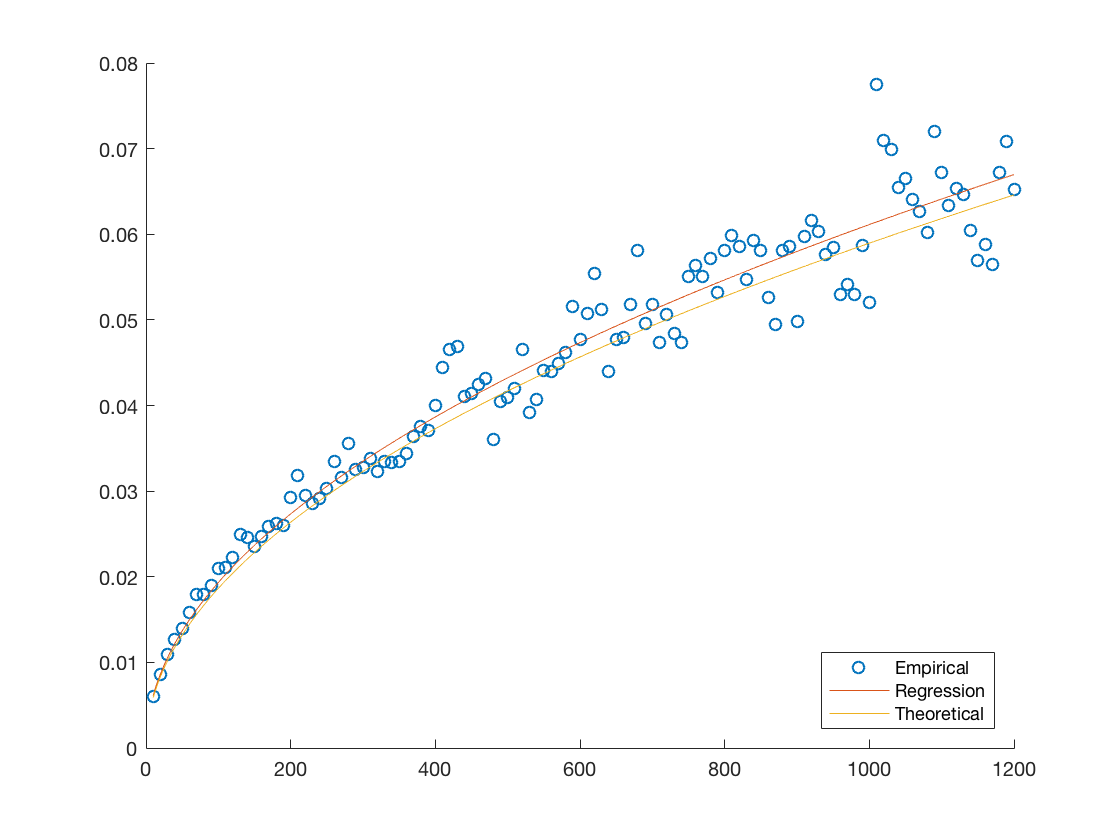}
\end{subfigure}

\vspace{3mm}

\begin{subfigure}[t]{0.49\textwidth}
\captionsetup{labelformat=empty}

\caption{MSFT}
\includegraphics[width=\textwidth, trim = 0 0 0 30, clip]{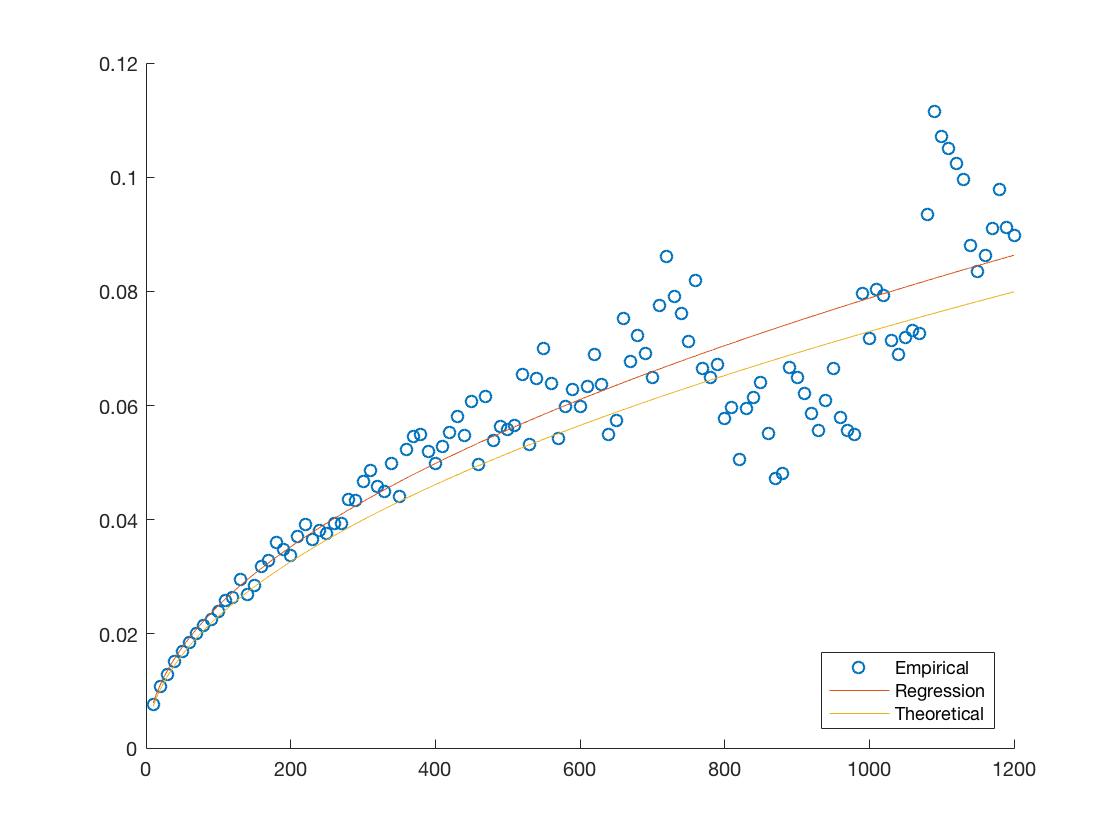}
\end{subfigure}

\caption{\label{fig:bestfit} A qualitative comparison of the regression to the theoretical model. For APPL, AMZN and GOOG we have used a Markov chain generated from 16 quantiles taken on the upward movements and downward movements. For INTC and MSFT we have taken the CHPDO coefficient since a different coefficient is not possible with the other models.}

\end{figure}

Another quantitative measure of our the fit would be to compare them to the best theoretical one available given the data. Notice that each of our models assumes that the standard deviation evolves according to the square root of the time step times some determinable coefficient. Therefore we can estimate the best possible coefficient by minimizing the mean squared residual, we provide plots of these hypothetical best fits against the empirical data and theoretical fits in Figure \ref{fig:bestfit}. We also provide the coefficients for the theoretical fits and hypothetical best fit in Table \ref{table:bestfit} in order to have a more quantitative comparison.

\begin{table}[htpb]
    \centering
    \begin{tabular}{c|c|c|c}
     & Theoretical Coefficient & Regression Coefficient & Percent Error\\
    \hline
    AAPL & 0.02868 & 0.02828 & 1.42\% \\
    AMZN & 0.01450 & 0.01831 & 20.8\% \\
    GOOG & 0.02883 & 0.03023 & 4.63\% \\
    INTC & 0.00186 & 0.00193 & 3.4\% \\
    MSFT & 0.00231 & 0.00246 & 6.4\%
    \end{tabular}
    \caption{\label{table:bestfit} The coefficients calculated for AAPL, AMZN and GOOG are generated using a Markov chain created by 16 quantiles on the upward and downward movements. While the coefficients for INTC and MSFT are obtained from the CHPDO case.}
    
\end{table}

We notice that in each case the errors are close to, or under five percent with AMZN being the biggest offender. This is consistent with the discussion provided throughout our analysis and highlights the general applicability of our model.

\subsection{Remarks}

Overall the N state model out performs the others, generating fits for four out of the five datasets that are reasonable and providing reductions in the mean squared error by upwards of 25\%. While not able to capture the full dynamics observed in AMZN it appears to be a strong candidate for a simple model of price dynamics observed in our data. Further investigation would be necessary to determine what causes the additional volatility observed in AMZN, and potentially implement a more robust model which captures this.


\end{document}